\documentclass[aps, prd, amsmath, floats, floatfix, twocolumn, nofootinbib, superscriptaddress]{revtex4-1}
\usepackage[pdftex]{graphicx}
\usepackage{color}
\usepackage{latexsym}
\usepackage[colorlinks]{hyperref}

\newcommand{\be}{\begin{eqnarray}}
\newcommand{\ee}{\end{eqnarray}}

\begin{document}

\title{Testing the Kerr hypothesis using X-ray reflection spectroscopy with \textsl{NuSTAR} data of Cygnus~X-1 in the soft state}

\author{Honghui~Liu}
\affiliation{Center for Field Theory and Particle Physics and Department of Physics, Fudan University, 200438 Shanghai, China}

\author{Askar~B.~Abdikamalov}
\affiliation{Center for Field Theory and Particle Physics and Department of Physics, Fudan University, 200438 Shanghai, China}

\author{Dimitry~Ayzenberg}
\affiliation{Center for Field Theory and Particle Physics and Department of Physics, Fudan University, 200438 Shanghai, China}

\author{Cosimo~Bambi}
\email[Corresponding author: ]{bambi@fudan.edu.cn}
\affiliation{Center for Field Theory and Particle Physics and Department of Physics, Fudan University, 200438 Shanghai, China}

\author{Thomas~Dauser}
\affiliation{Remeis Observatory \& ECAP, Universit\"{a}t Erlangen-N\"{u}rnberg, 96049 Bamberg, Germany}

\author{Javier~A.~Garc{\'\i}a}
\affiliation{Cahill Center for Astronomy and Astrophysics, California Institute of Technology, Pasadena, CA 91125, USA}
\affiliation{Remeis Observatory \& ECAP, Universit\"{a}t Erlangen-N\"{u}rnberg, 96049 Bamberg, Germany}

\author{Sourabh~Nampalliwar}
\affiliation{Theoretical Astrophysics, Eberhard-Karls Universit\"at T\"ubingen, 72076 T\"ubingen, Germany}

\begin{abstract}
We continue exploring the constraining capabilities of X-ray reflection spectroscopy to test the Kerr-nature of astrophysical black holes and we present the results of our analysis of two \textsl{NuSTAR} observations of Cygnus~X-1 in the soft state. We find that the final measurement can strongly depend on the assumption of the intensity profile. We conclude that Cygnus~X-1 is not suitable for tests of general relativity using X-ray reflection spectroscopy and we discuss the desired properties of a source to be a good candidate for our studies.
\end{abstract}

\maketitle


\section{Introduction}

The theory of general relativity (GR) was proposed over a century ago by Einstein and is still one of the pillars of modern physics. It has been extensively tested in the weak field regime, mainly with experiments in the Solar System and accurate radio observations of binary pulsars, and current data are in agreement with the theoretical predictions~\cite{Will:2014kxa}. However, the validity of GR in the strong field regime is still largely unexplored, and there are a number of gravitational theories that have the same predictions as GR in weak gravitational fields and present deviations from GR when gravity becomes strong. Gravitational tests in the strong field regime are thus crucial to distinguish GR from these alternative models and are becoming a hot topic today.

Black holes are ideal laboratories for GR tests in the strong field regime. In 4-dimensional Einstein's gravity, the spacetime of a rotating, uncharged black hole is described by the Kerr metric~\cite{h1,h2,h3} and is completely characterized by only two parameters, which represent, respectively, the mass $M$ and the spin angular momentum $J$ of the compact object. The spacetime around an astrophysical black hole should be very well approximated by the ideal Kerr solution of GR. Once the black hole is formed, initial deviations from the Kerr metric should be quickly radiated away by the emission of gravitational waves~\cite{price}. Deviations from the Kerr solution due to the presence of accretion disks~\cite{k1}, nearby stars~\cite{k2}, or due to a possible non-vanishing electric charge of the black hole~\cite{k3}, are normally extremely small and can be safely ignored. Testing the Kerr-nature of astrophysical black holes is thus a GR test in the strong field regime~\cite{t1,t2,t3,t4,t5,t6-0,t6,t7,t8} and can be seen as the natural evolution of the GR tests in the weak field regime with Solar System experiments.

X-ray reflection spectroscopy refers to the analysis of the reflection spectrum of accretion disks around black holes~\cite{x1,x2,x3}. The reflection spectrum is the result of the illumination of the disk by the so-called ``corona'', which is a hot cloud close to the black hole, even if its exact morphology is not yet well understood. The reflection spectrum is characterized by some fluorescent emission lines, in particular the iron K$\alpha$ complex at 6.4-7~keV, and the Compton hump at 20-30~keV. The reflection spectrum in the rest-frame of the gas in the disk can be derived from atomic physics calculations. The reflection spectrum that we observe is the result of relativistic effects (Doppler boosting, gravitational redshift, light bending) occurring in the strong gravity region near the black hole. The study of the features in the reflection spectrum can be used to determine the spacetime geometry of the black hole\footnote{We note that the relativistic origin of broad iron lines was under debate for a while, because these broad iron lines could also be explained by models with warm absorbers~\cite{abs1,abs2}. However, there is today a common consensus on their relativistic origin. After the launch of \textsl{NuSTAR} in 2012, data include the Compton hump at 20-30~keV, which cannot be explained by absorber models~\cite{risaliti13}. Reverberation measurements are also consistent with a reflection spectrum generated from the inner part of the disk and cannot be explained within absorber models~\cite{uttley14}.}.

Recently, our group has extended the relativistic reflection model {\sc relxill}~\cite{j1,j2,j3} to parametric black hole spacetimes. The new model, which we called {\sc relxill\_nk}~\cite{noi1,noi2}, is designed to test the Kerr nature of astrophysical black holes. {\sc relxill\_nk} calculates reflection spectra in a parametric black hole spacetime, where some ``deformation parameters'' quantify certain deformations from the Kerr background. From the comparison of the theoretical predictions of {\sc relxill\_nk} with observational data of specific sources, we can constrain the values of these deformation parameters and check whether they vanish, as it would be required in GR. Current constraints on possible non-Kerr features in the spacetime metric around specific sources have been reported in our previous work~\cite{noi3,noi4,noi5,noi6,noi7,noi8,noi9,noi10,noi11}.

In the present paper, we continue exploring the capabilities of X-ray reflection spectroscopy to test the Kerr black hole hypothesis and we extend our study to two \textsl{NuSTAR} observations of the stellar-mass black hole in Cygnus~X-1 in the soft state. Cygnus~X-1 is a very bright source, several studies in literature have found that the value of its spin parameter is quite high, and the quality of the \textsl{NuSTAR} data is good. Despite that, we find that it is difficult to test the Kerr nature of this object because of problems in modeling its spectrum. The final measurement depends on the assumptions about the intensity profile of the reflection component and in some cases we do not recover the Kerr solution of GR. The uncertainties in the estimate of the deformation parameters are also large in comparison with other sources. Similar problems were partially found in our previous test with the stellar-mass black hole in GRS~1915+105~\cite{noi10}, while we have never met these issues in the tests with supermassive black holes. On the basis of all these studies, we discuss which properties a source and the observations should have for our tests of the Kerr metric using X-ray reflection spectroscopy. Supermassive black holes seems to be more suitable for our tests, probably because their spectrum is easier to model (the disk temperature is much lower), their variability time scales are longer, and they naturally have spin parameters very close to 1 (while there are only a few stellar-mass black holes with a possible high spin and those sources turn out to have complicated spectra).

The manuscript is organized as follows. In Section~\ref{s-metric}, we review the Johannsen metric~\cite{tj}, which is the parametric black hole background employed in our study. In Section~\ref{s-red}, we present the two \textsl{NuSTAR} observations of our analysis and we briefly describe the data reduction. In Section~\ref{s-ana}, we present the spectral analysis and the constraints on the deformation parameters. Section~\ref{s-dis} is devoted to the discussion of our results and Section~\ref{s-con} is for the conclusions. Throughout the paper, we adopt units in which $G_{\rm N} = c = 1$ and a metric with signature $(-+++)$.


\section{Parametric black hole spacetime \label{s-metric}}

Parametric black hole spacetimes are a common choice for tests of the Kerr hypothesis with electromagnetic techniques. The Kerr metric is deformed by adding some deformation parameters, which are introduced to quantify deviations from the Kerr solution. The properties of the electromagnetic spectrum are calculated in this more general spacetime, where the deformation parameters are just some of the parameters of the whole model. From the comparison of the theoretical predictions with observations, we can infer the values of these deformation parameters and verify if they indeed vanish, as it is necessary in order to recover the Kerr solution. While the nature of this deformed metric is questionable, in the end the spirit is the same as in a null experiment. We expect that the deformation parameters vanish and we want to measure that this is indeed the case. If an astrophysical measurement required that at least one of the deformation parameters is non-vanishing, then the spacetime metric of the black hole would not be described by the Kerr solution, but it would not be possible to determine the exact form of the spacetime metric with these tests.

In this paper, we employ the Johannsen metric~\cite{tj}. In Boyer-Lindquist-like coordinates, the line element reads
\be\label{eq-jm}
ds^2 &=&-\frac{\tilde{\Sigma}\left(\Delta-a^2A_2^2\sin^2\theta\right)}{B^2}dt^2
+\frac{\tilde{\Sigma}}{\Delta A_5}dr^2+\tilde{\Sigma} d\theta^2 \nonumber\\
&&-\frac{2a\left[\left(r^2+a^2\right)A_1A_2-\Delta\right]\tilde{\Sigma}\sin^2\theta}{B^2}dtd\phi \nonumber\\
&&+\frac{\left[\left(r^2+a^2\right)^2A_1^2-a^2\Delta\sin^2\theta\right]\tilde{\Sigma}\sin^2\theta}{B^2}d\phi^2 \, ,
\ee
where $M$ is the black hole mass, $a = J/M$, $J$ is the black hole spin angular momentum, $\tilde{\Sigma} = \Sigma + f$, and
\be
\Sigma &=& r^2 + a^2 \cos^2\theta \, , \\
\Delta &=& r^2 - 2 M r + a^2 \, , \\
B &=& \left(r^2+a^2\right)A_1-a^2A_2\sin^2\theta \, .
\ee
The functions $f$, $A_1$, $A_2$, and $A_5$ are defined as
\be
f &=& \sum^\infty_{n=3} \epsilon_n \frac{M^n}{r^{n-2}} \, , \\
A_1 &=& 1 + \sum^\infty_{n=3} \alpha_{1n} \left(\frac{M}{r}\right)^n \, , \\
A_2 &=& 1 + \sum^\infty_{n=2} \alpha_{2n} \left(\frac{M}{r}\right)^n \, , \\
A_5 &=& 1 + \sum^\infty_{n=2} \alpha_{5n} \left(\frac{M}{r}\right)^n \, ,
\ee
where $\{ \epsilon_n \}$, $\{ \alpha_{1n} \}$, $\{ \alpha_{2n} \}$, and $\{ \alpha_{5n} \}$ are four infinite sets of deformation parameters. This form of the Johannsen metric recovers the correct Newtonian limit and its deformation parameters are not constrained by Solar System experiments.

In what follows, we will restrict our analysis to the deformation parameters $\alpha_{13}$ and $\alpha_{22}$. These are indeed the two deformation parameters with the strongest impact on the reflection spectrum of the accretion disk~\cite{noi1}. The Kerr solution is recovered when $\alpha_{13} = \alpha_{22} = 0$. Here, we will only consider the possibility that one of the deformation parameters is non-vanishing; that is, we will measure $\alpha_{13}$ assuming $\alpha_{22} = 0$ and then we will measure $\alpha_{22}$ assuming $\alpha_{13} = 0$. The current version of {\sc relxill\_nk} does not allow for two free deformation parameters at the same time.

The parameters of the spacetime metric in our tests are the spin parameter $a_*= a/M = J/M^2$ and the deformation parameters $\alpha_{13}$ and $\alpha_{22}$. These parameters cannot have arbitrary values or, otherwise, we can have spacetimes with pathological properties (spacetime singularities, regions with closed time-like curves, etc.). As in the Kerr metric, the condition on the spin parameter is
\be
- 1 < a_* < 1 \, .
\ee 
For $| a_* | > 1$ there is no horizon and thus no black hole, and the central singularity is naked. For the deformation parameters $\alpha_{13}$ and $\alpha_{22}$, we impose the following conditions
\be
\label{eq-constraints}
&& \alpha_{13} > - \frac{1}{2} \left( 1 + \sqrt{1 - a^2_*} \right)^4 \, , 
\nonumber\\
&& - \left(1 + \sqrt{1 - a_*^2} \right)^2 < \alpha_{22} < \frac{\left( 1 + \sqrt{1 - a^2_*} \right)^4}{a_*^2} \, ,
\ee
which are obtained by imposing no violation of Lorentzian signature, no closed time-like curves, and no divergences in the metric on and outside the black hole event horizon (see Refs.~\cite{tj,noi4} for the details).


\begin{table}[t]
\centering
\vspace{0.5cm}
\begin{tabular}{cccc}
\hline\hline
Epoch & \hspace{0.1cm} Obs.~ID \hspace{0.1cm} & \hspace{0.1cm} Start date \hspace{0.1cm} & \hspace{0.1cm} Exposure (ks) \hspace{0.1cm} \\
\hline\hline
1 & 00001011001 & 2012-07-02 & 14.4 \\
& 00001011002 & & 5.2 \\
\hline
2 & 10002003001 & 2012-07-06 & 12.5 \\
\hline
3 & 30001011002 & 2012-10-31 & 11.0 \\
& 30001011003 & & 5.7 \\
& 10014001001 & & 4.6 \\
\hline
4 & 30001011009 & 2014-10-04 & 22.6 \\
\hline\hline
\end{tabular}
\vspace{0.2cm}
\caption{\textsl{NuSTAR} observations of Cygnus~X-1 in the soft state analyzed in~\cite{w16}. In our paper, we study the observations of epoch~1 and 4. \label{t-obs}}
\end{table}

\section{Observations and data reduction \label{s-red}}

Cygnus X-1 is a very bright source in the sky. It was one of the first detected X-ray sources and also the first dynamically confirmed black hole. It is a high mass X-ray binary and the accreting gas around the black hole comes from the wind of the companion star. The system is at a distance $1.86^{+0.12}_{-0.11}$~kpc~\cite{Reid:2011nn} and the mass of the black hole is estimated to be $14.8 \pm 1.0$~$M_{\odot}$~\cite{Orosz:2011np}.

Cygnus~X-1 is quite a popular source and has been studied by many authors. There are several analyses in the literature of the reflection spectrum of this source and with data from different X-ray missions~\cite{c1,c2,c3,c4,c5,w16,c6}. All the previous studies assumed the Kerr metric and found a high value of the spin parameter of the black hole (say, $a_* > 0.9$) at a high confidence level. These measurements are confirmed by the analysis of the thermal spectrum of the disk (continuum-fitting method): still assuming the Kerr metric, in Refs.~\cite{lijun1,lijun2} the authors find that the spin parameter of the black hole in Cygnus~X-1 may be close to 1.

In the present work, we consider the observations of \textsl{NuSTAR}~\cite{nustar} analyzed in~\cite{w16} and summarized in Tab.~\ref{t-obs}. In what follows, we only discuss the observations of epoch~1 and 4, because those of epoch~2 and 3 are characterized by strong absorption by the stellar wind of the companion star and are definitively less suitable for tests of the Kerr metric.

We reduce the data from instruments Focal Plane Modules A and B (FPMA and FPMB). For epoch~1, we separately reduced the data of the two observations and then we used ADDASCASPEC to combine the spectra for each FPM instrument. Version 0.4.6 of \textit{nupipeline} and version 20180419 of \textsl{NuSTAR} Calibration Database (CALDB) were used to produce cleaned event files. We ran \textit{nuproduct} to extract light curves, spectra, and response files. The source was extracted with a circular region centered on Cygnus~X-1 with a radius of 150''. The background region was a circle with the same size taken farthest from the source region to avoid contribution from the source. The spectra were then grouped to have a minimal count of 50 for each bin in order to use $\chi^2$-statistics.


\section{Spectral analysis \label{s-ana}}

For the spectral analysis, we employ XSPEC v12.10.0c~\cite{xspec}. As in Ref.~\cite{w16}, we analyze the spectra with the XSPEC model

\begin{widetext}
{\sc tbabs $\times$ xstar $\times$ (diskbb + cutoffpl + gaussian + relconv\_nk $\times$ gsmooth $\times$ xillver)}
\end{widetext}

Let us now briefly describe every component:
\begin{enumerate}
\item {\sc tbabs} -- It describes the Galactic absorption and has one parameter: the column density $N_{\rm H}$. We use the abundances of Wilms et al.~\cite{wilms-2000} and the cross sections of Verner et al.~\cite{Verner-1996}. The column density $N_{\rm H}$ is fixed to $6.0 \times 10^{21}$~cm$^2$. 
\item {\sc diskbb}~\cite{diskbb} -- It is a multi-temperature blackbody model and accounts for the thermal component from the accretion disk. It has two parameters: the inner temperature of the disk, $T_{\rm in}$, and the normalization.
\item {\sc cutoffpl} -- For the power-law component from the corona, we employ {\sc cutoffpl} which is a power-law with an exponential cut-off energy $E_{\rm cut}$. The model has three parameters: the photon index, $\Gamma$, the cut-off energy, $E_{\rm cut}$, and the normalization.
\item {\sc xillver} -- The power-law component illuminating the disk produces the reflection spectrum, here described by {\sc xillver}~\cite{xillver}. In {\sc xillver}, the values of $\Gamma$ and $E_{\rm cut}$ are tied to those in {\sc cutoffpl} and the reflection fraction is set to $-1$ so it returns the reflection component only. The model has thus three parameters: the iron abundance, $A_{\rm Fe}$, the ionization parameter, $\xi$, and the normalization.
\item {\sc relconv\_nk}~\cite{noi2} -- It is our convolution model to take all the relativistic effects (Doppler boosting, gravitational redshift, light bending) in the Johannsen spacetime into account assuming a Novikov-Thorne accretion disk. There are seven parameters: the spin parameter $a_*$, the inclination angle of the disk with respect to the line of sight of the distant observer, $i$, the deformation parameters $\alpha_{13}$ and $\alpha_{22}$, and three more parameters related to the intensity profile of the reflection component in the disk. The latter is described by a broken power-law, so we have the inner emissivity index $q_{\rm in}$, the outer emissivity index $q_{\rm out}$, and the breaking radius $R_{\rm br}$. In our study, we try three models for the intensity profile: 
\begin{itemize}
\item a simple power-law and we set $q_{\rm in} = q_{\rm out}$;
\item a broken power-law with $q_{\rm in}$ and $R_{\rm br}$ free and $q_{\rm out} = 3$ (lamppost coronal geometry);
\item a broken power-law with $q_{\rm in}$, $q_{\rm out}$, and $R_{\rm br}$ free.
\end{itemize}
\item {\sc gsmooth} -- Since {\sc xillver} assumes that the accretion disk is cold (the calculations ignore the radiation produced by the disk), while the inner part of accretion disks in X-ray binaries is 0.1-1~keV, the Compton broadening of emission lines is underestimated. We thus apply {\sc gsmooth} to {\sc xillver} before including the relativistic effects, as done in Ref.~\cite{w16}. 
\item {\sc gaussian} -- It describes a narrow neutral iron emission ($E=6.4$~keV) from the wind of the massive companion star.
\item {\sc xstar} -- The data show iron absorption features that, following Ref.~\cite{w16}, we interpret as ionized plasma in the stellar wind. We use the same {\sc xstar} model as in Ref.~\cite{w16}. The model has three parameters: the column density, $N_{\rm H}$, ionization parameter, $\xi$, and line-of-sight velocity of the wind, $v_{\rm out}$.
\end{enumerate}

\begin{figure*}[t]
\begin{center}
\includegraphics[width=8.5cm,trim={0.5cm 1.5cm 0.5cm 1.5cm},clip]{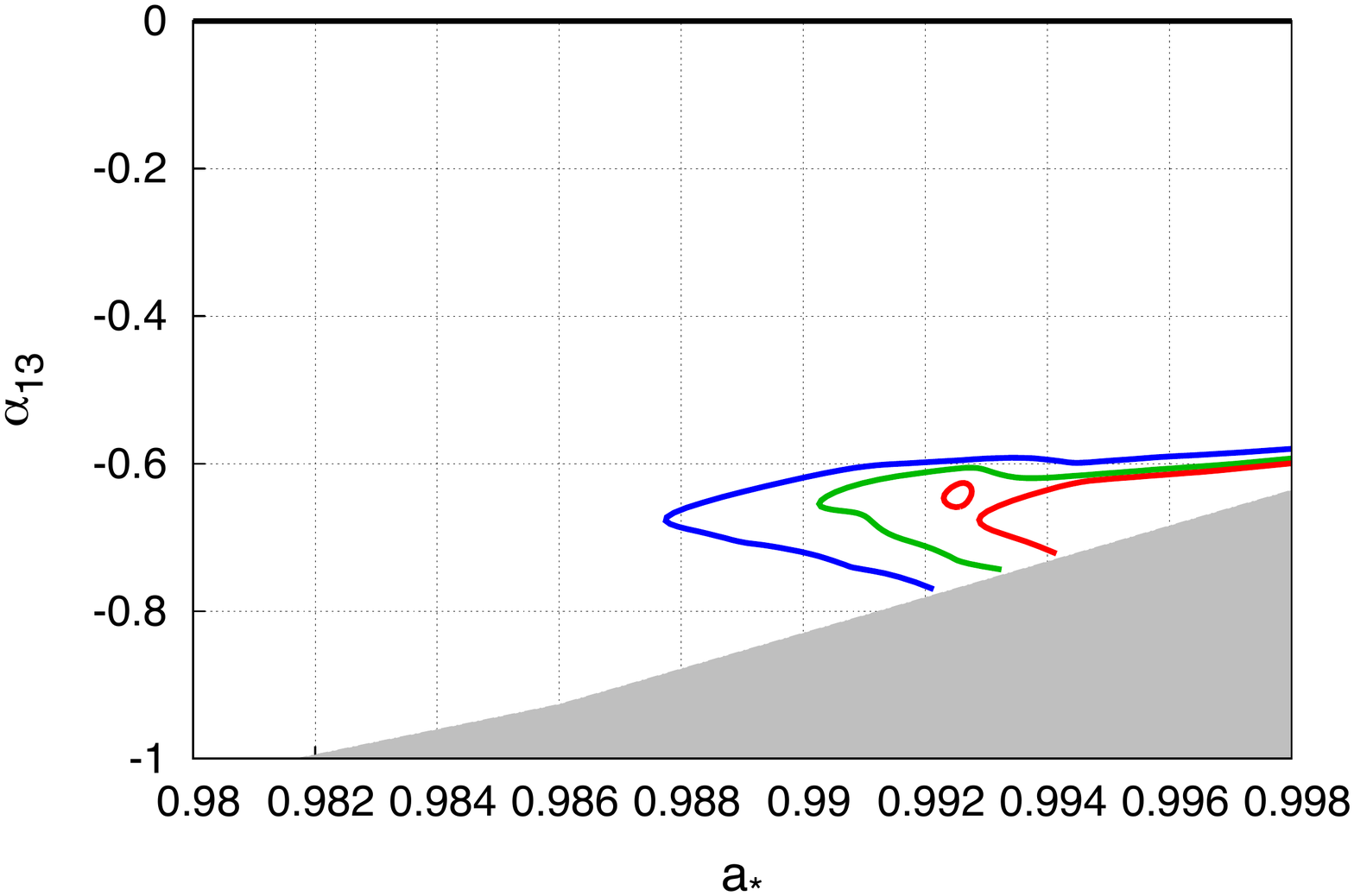}
\includegraphics[width=8.5cm,trim={0.5cm 1.5cm 0.5cm 1.5cm},clip]{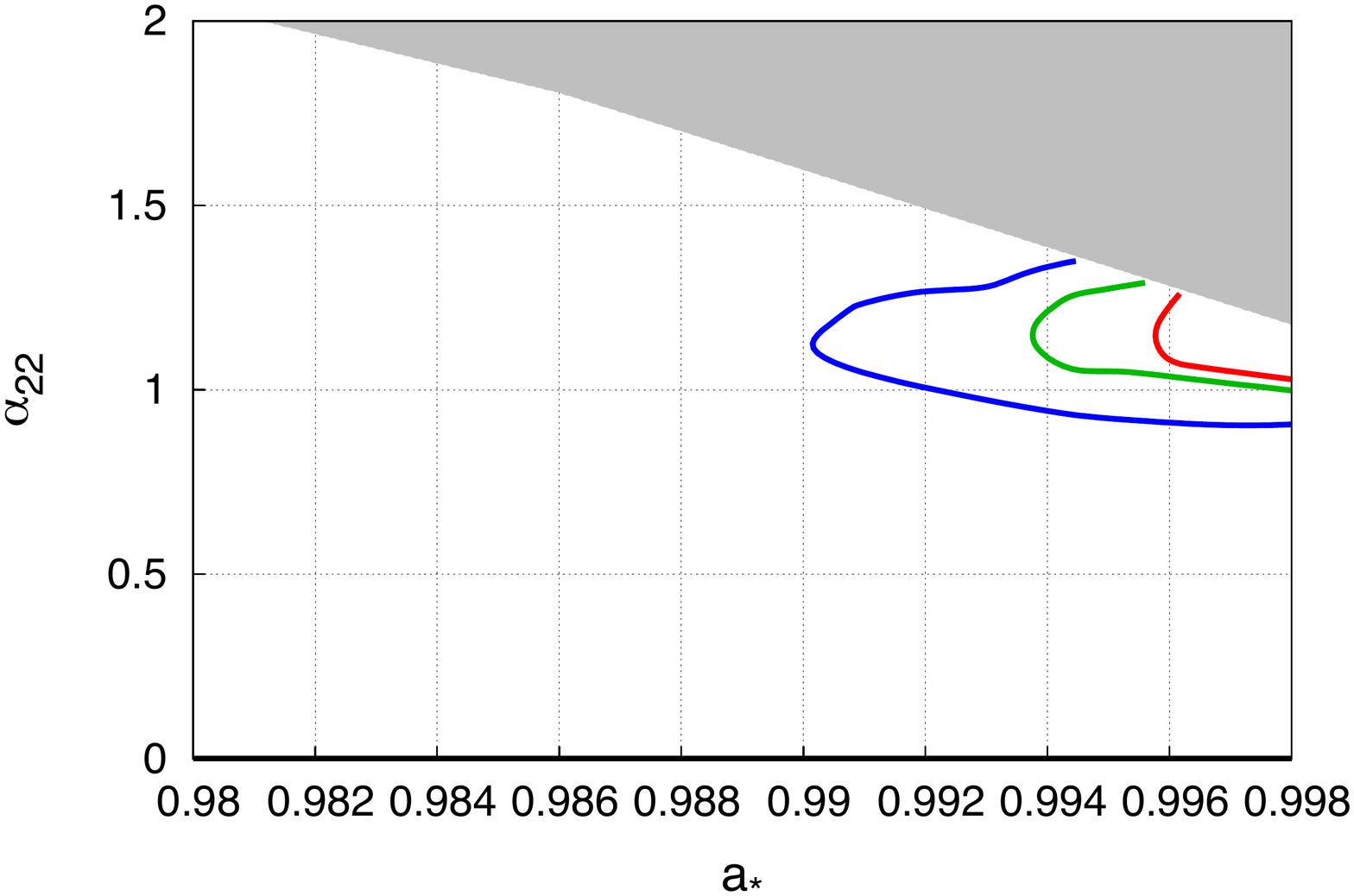} \\
\includegraphics[width=8.5cm,trim={0.5cm 1.5cm 0.5cm 1.5cm},clip]{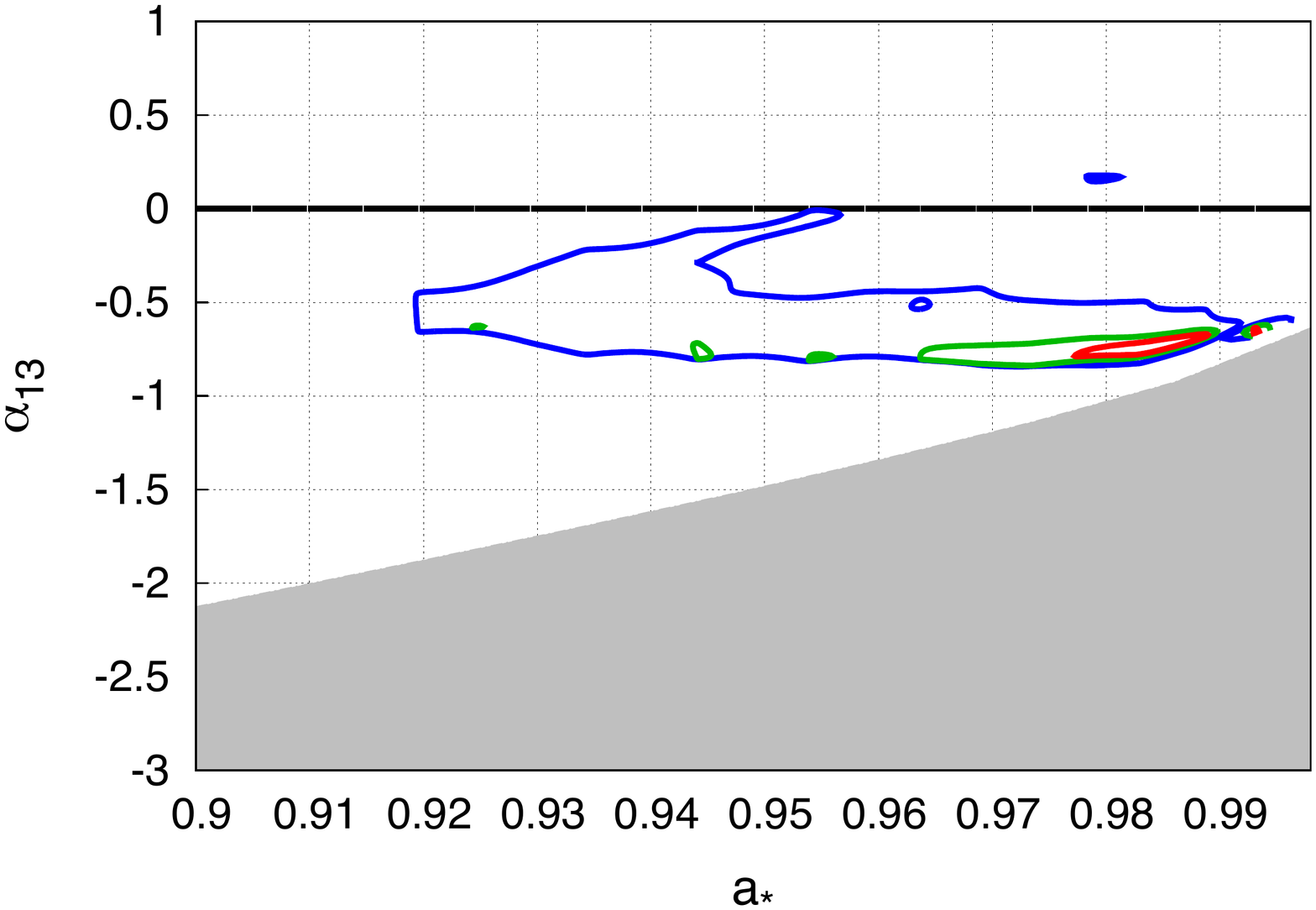}
\includegraphics[width=8.5cm,trim={0.5cm 1.5cm 0.5cm 1.5cm},clip]{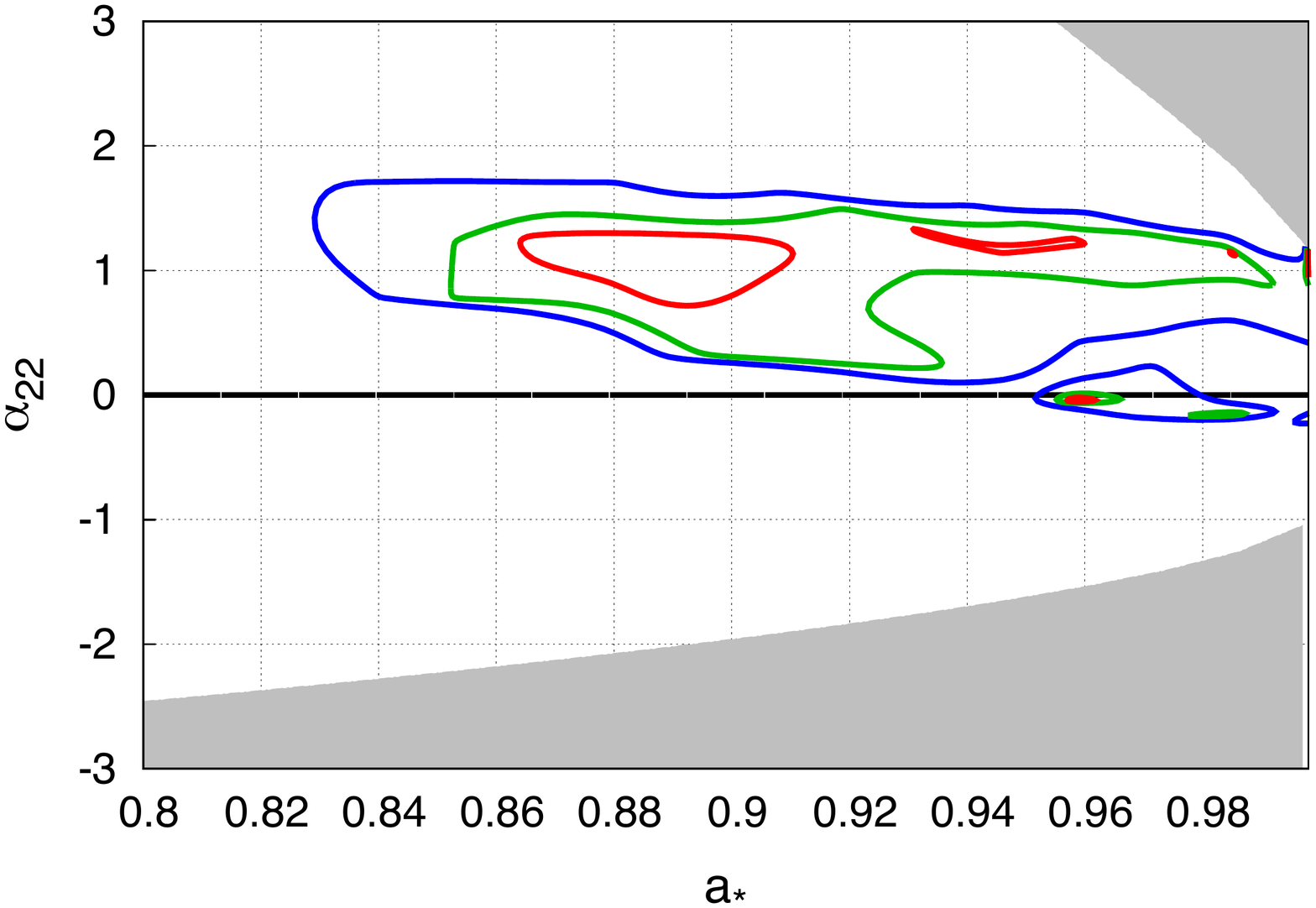} \\
\includegraphics[width=8.5cm,trim={0.5cm 1.5cm 0.5cm 1.5cm},clip]{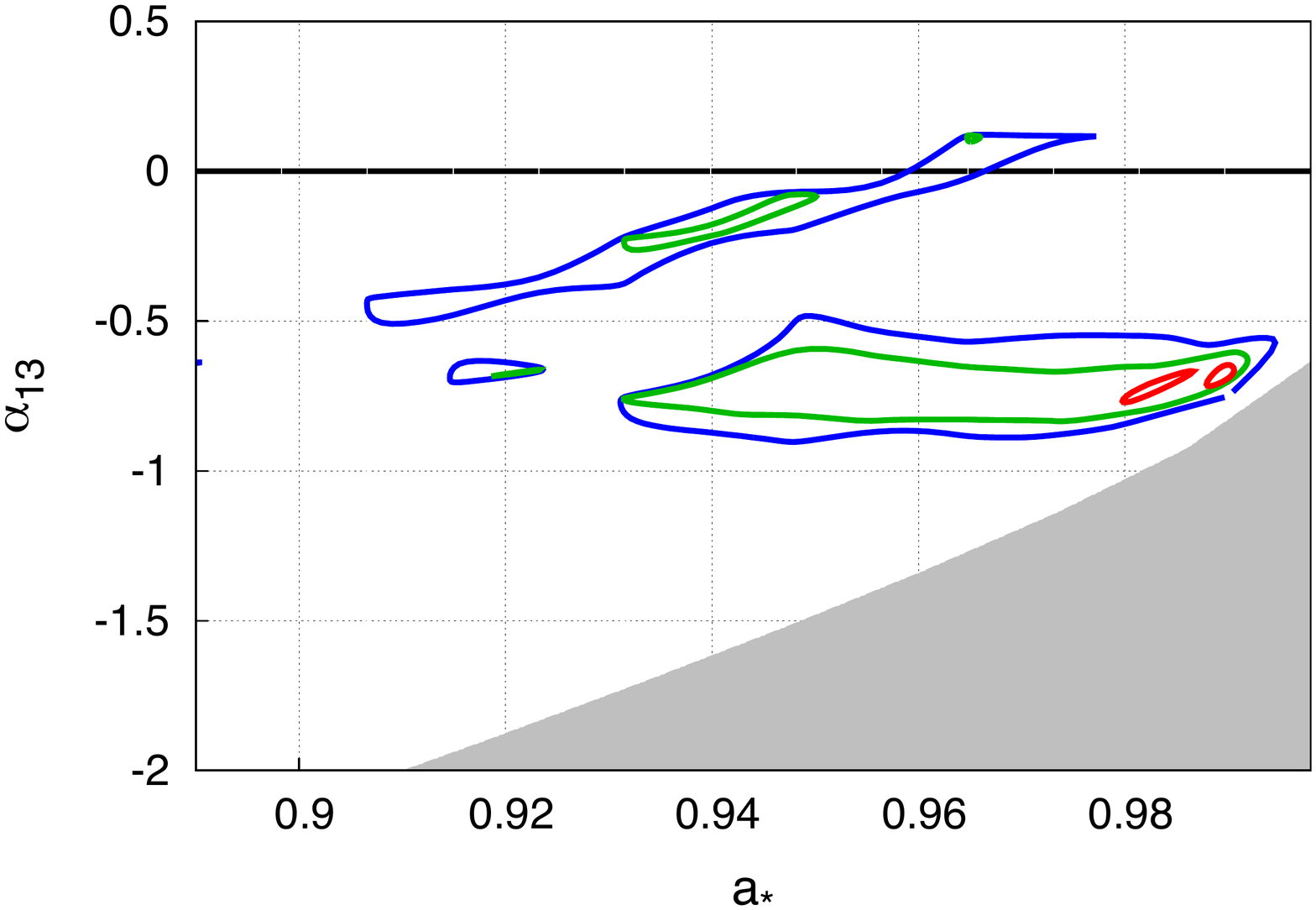}
\includegraphics[width=8.5cm,trim={0.5cm 1.5cm 0.5cm 1.5cm},clip]{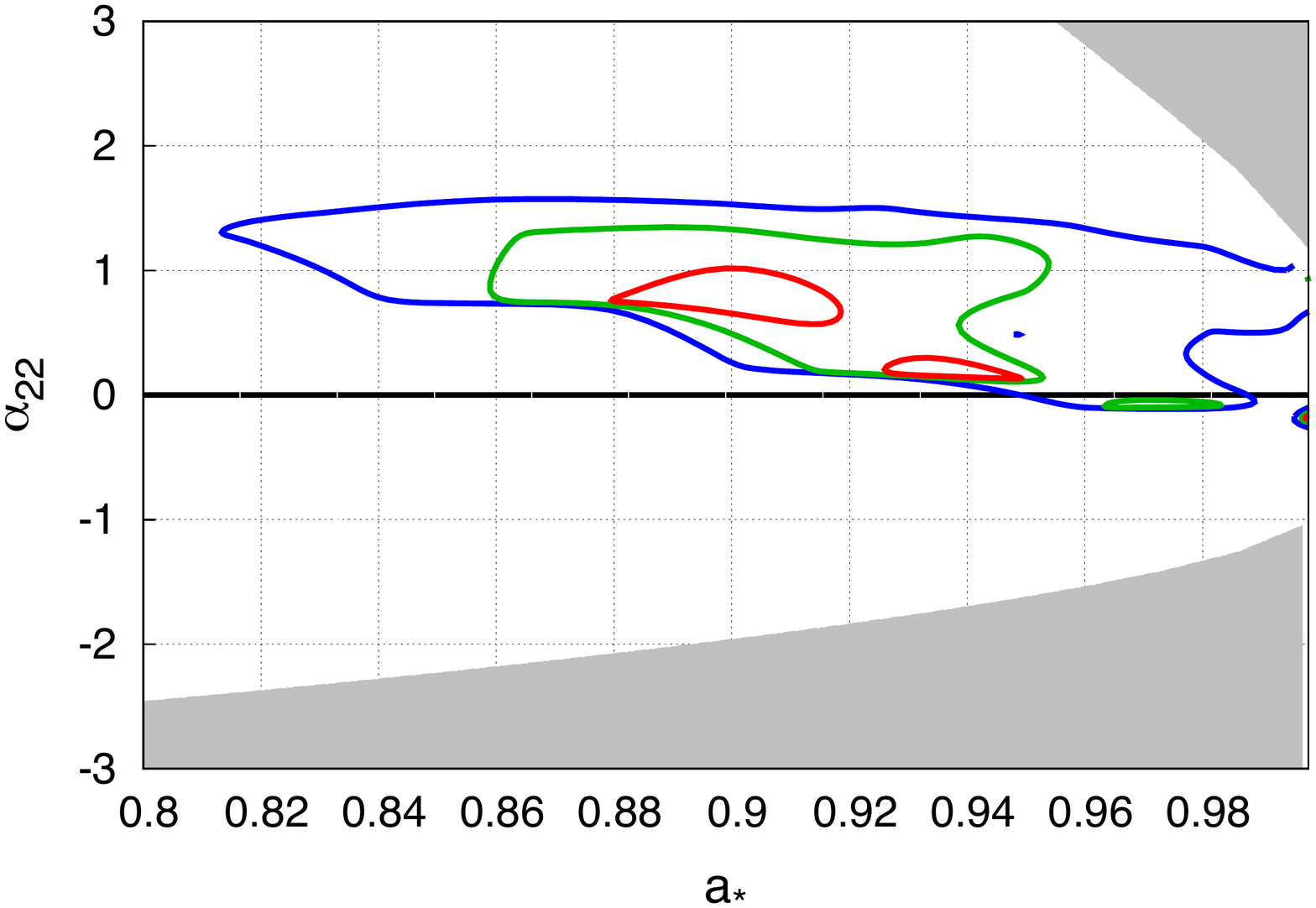}
\end{center}
\vspace{-0.6cm}
\caption{Constraints on the spin parameter $a_*$ and the Johannsen deformation parameters $\alpha_{13}$ (left panel) and $\alpha_{22}$ (right panel) from epoch~1. The emissivity profile is modeled with a power-law (top panels), a broken power-law with outer emissivity index frozen to 3 (central panels), and a broken power-law with both emissivity indices free (bottom panels). The red, green, and blue curves are, respectively, the 68\%, 90\%, and 99\% confidence level boundaries for two relevant parameters. \label{f-c1}}
\end{figure*}

\begin{figure*}[t]
\begin{center}
\includegraphics[width=8.5cm,trim={0.5cm 1.5cm 0.5cm 1.5cm},clip]{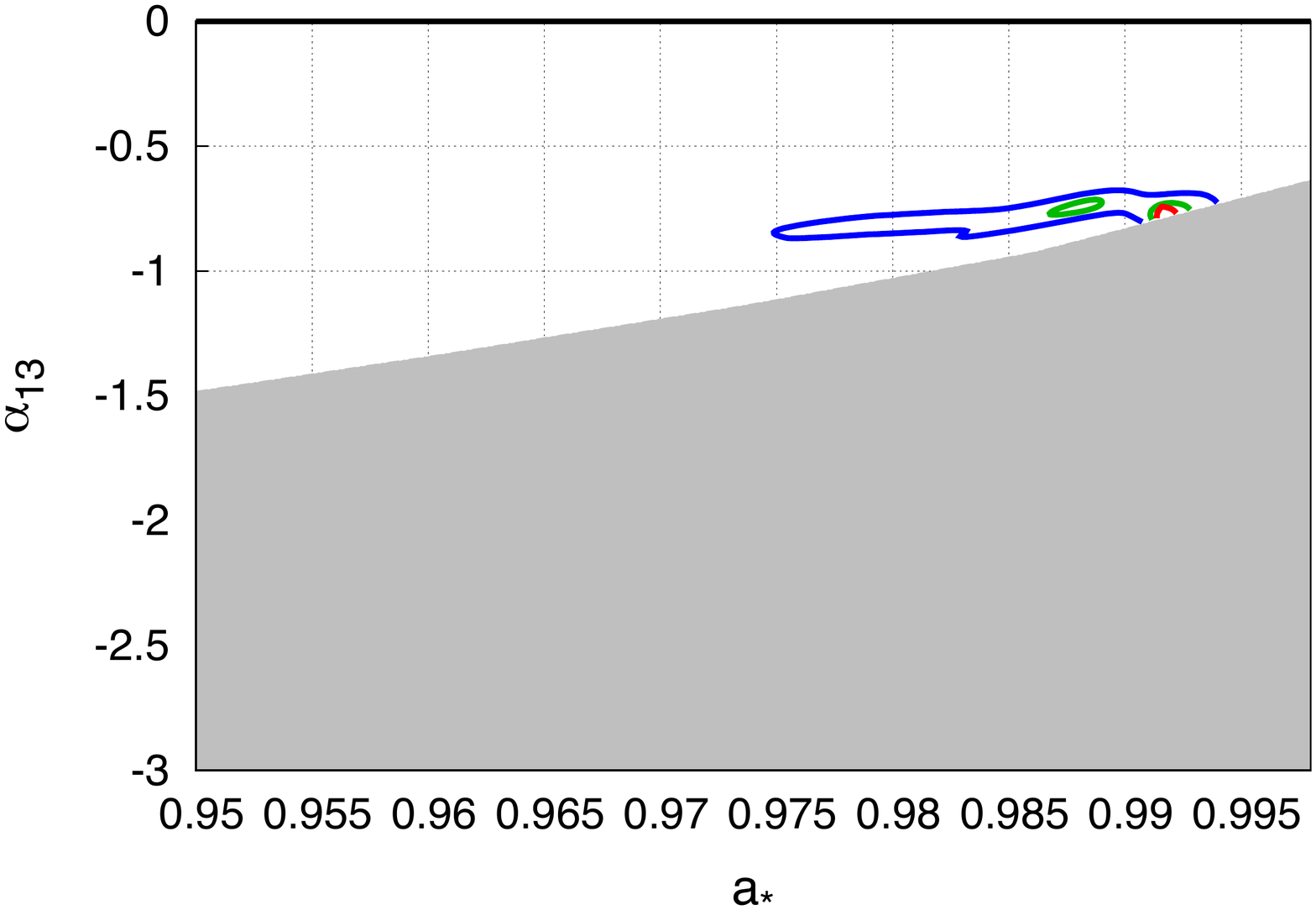}
\includegraphics[width=8.5cm,trim={0.5cm 1.5cm 0.5cm 1.5cm},clip]{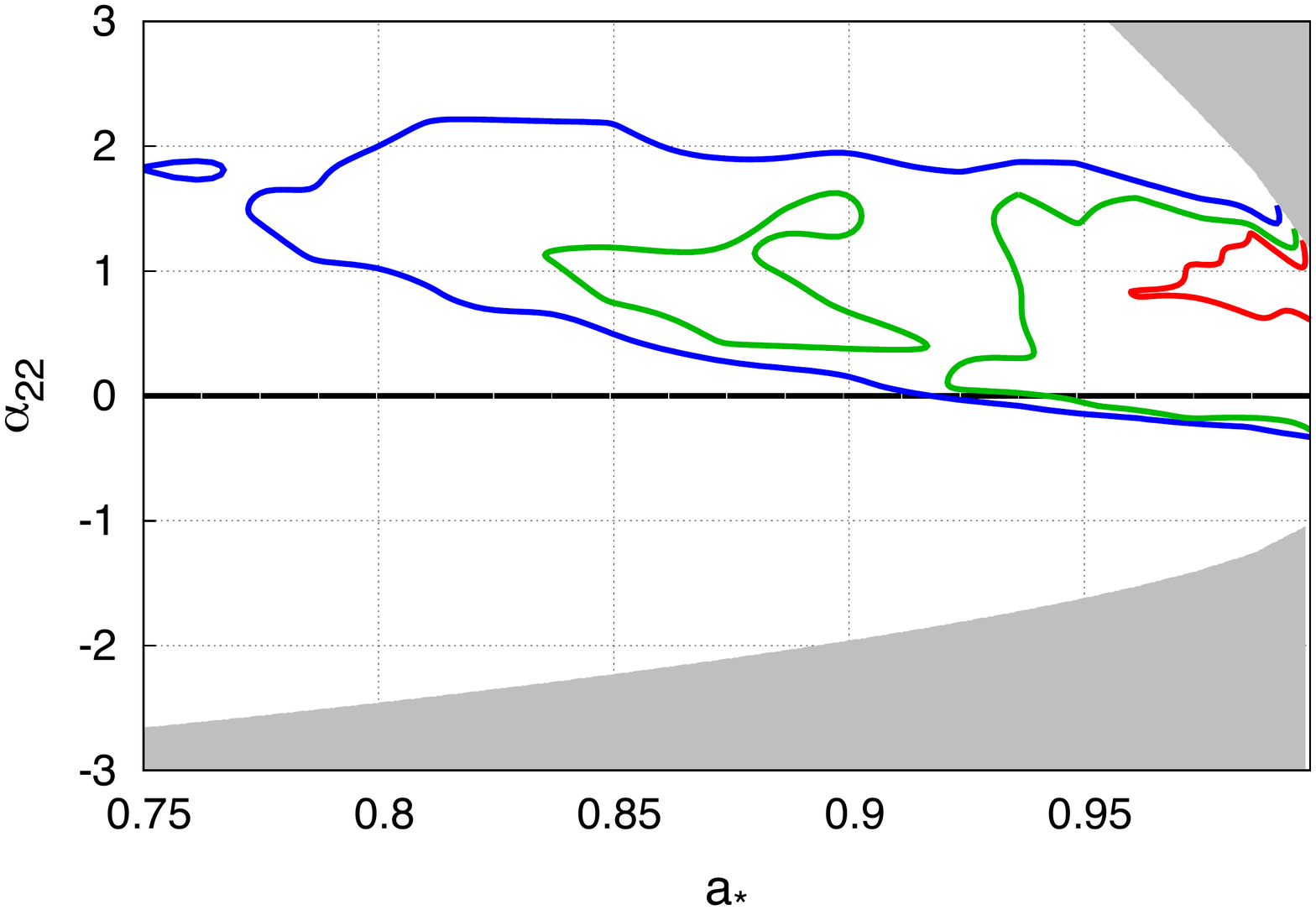} \\
\includegraphics[width=8.5cm,trim={0.5cm 1.5cm 0.5cm 1.5cm},clip]{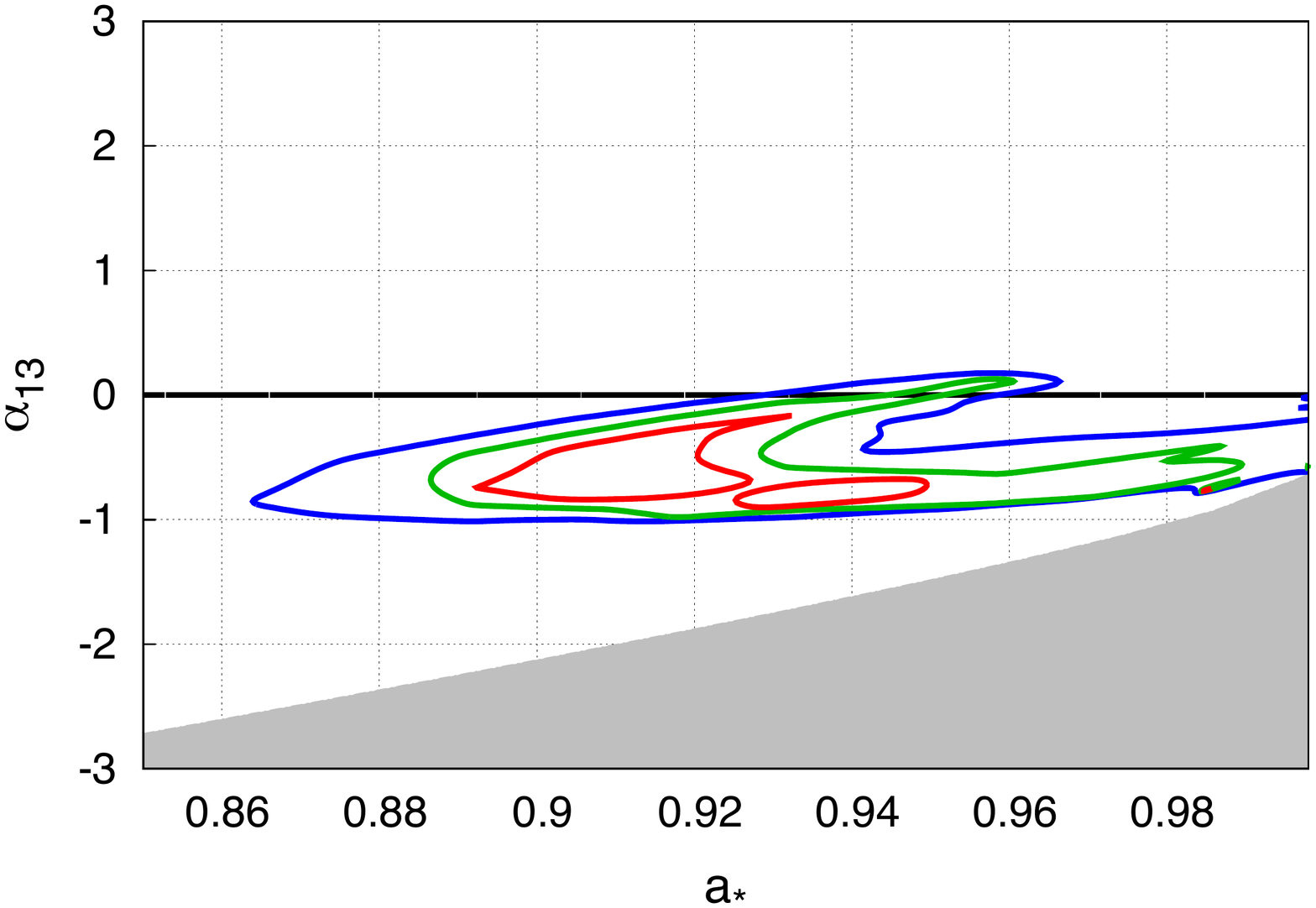}
\includegraphics[width=8.5cm,trim={0.5cm 1.5cm 0.5cm 1.5cm},clip]{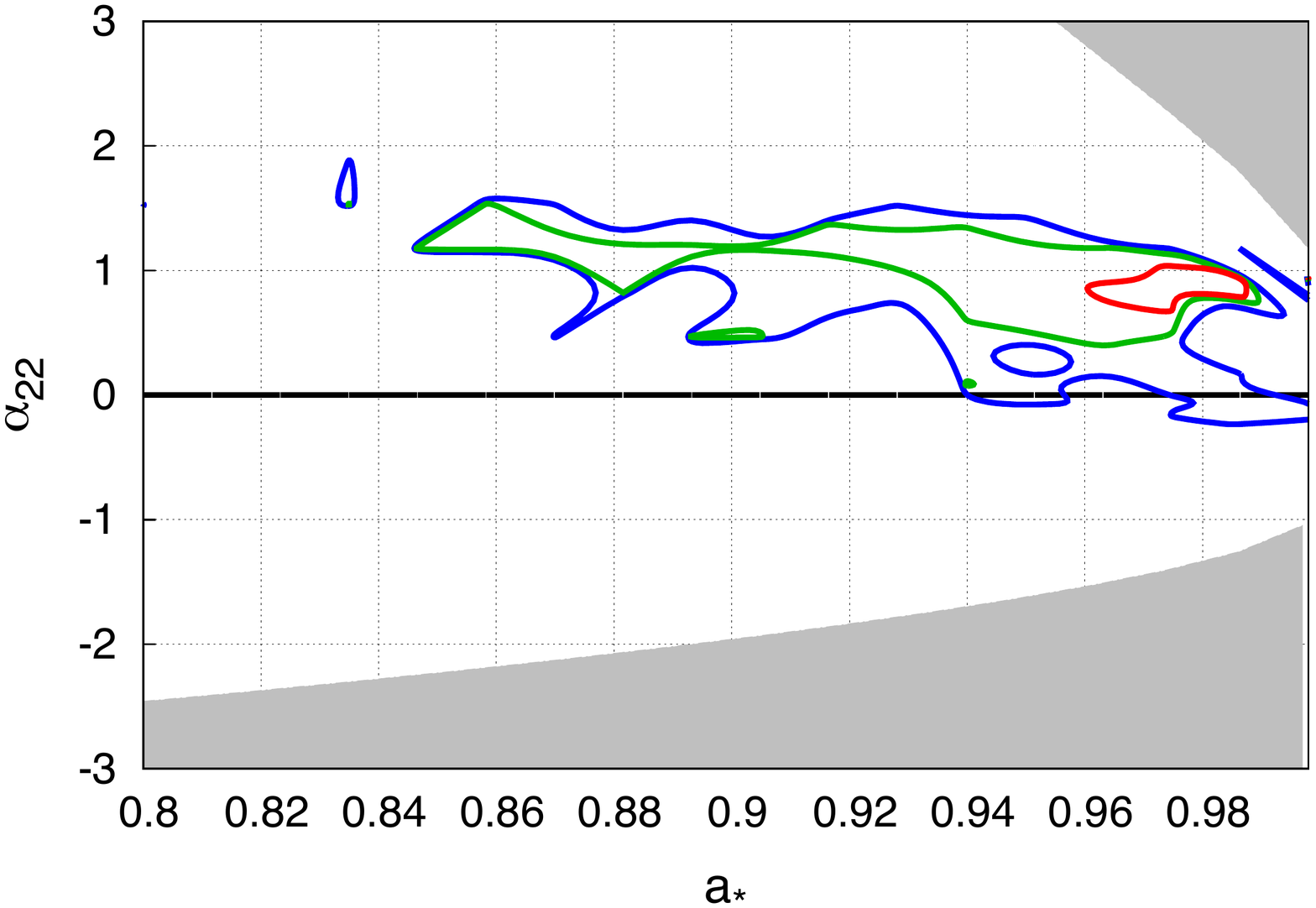} \\
\includegraphics[width=8.5cm,trim={0.5cm 1.5cm 0.5cm 1.5cm},clip]{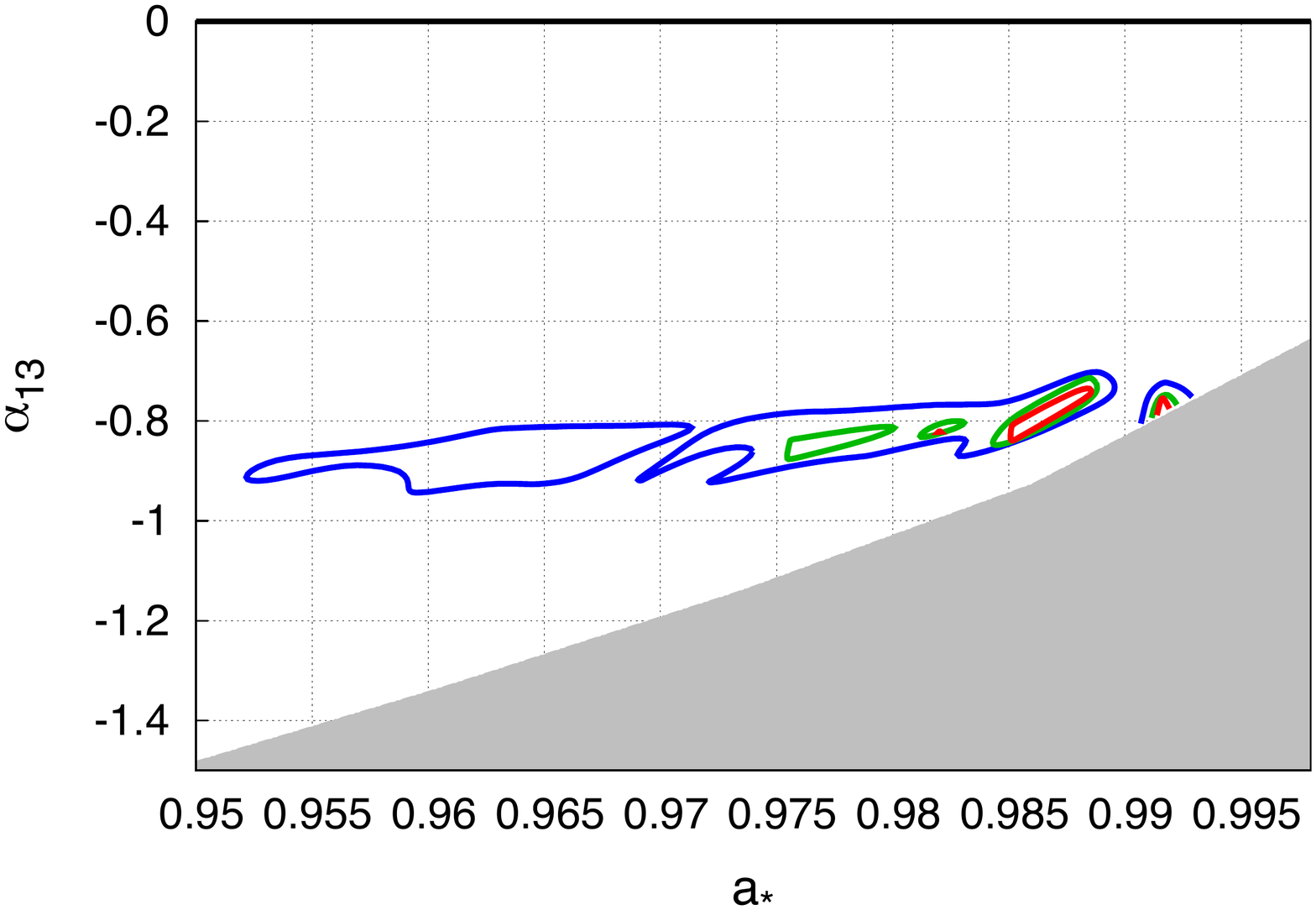}
\includegraphics[width=8.5cm,trim={0.5cm 1.5cm 0.5cm 1.5cm},clip]{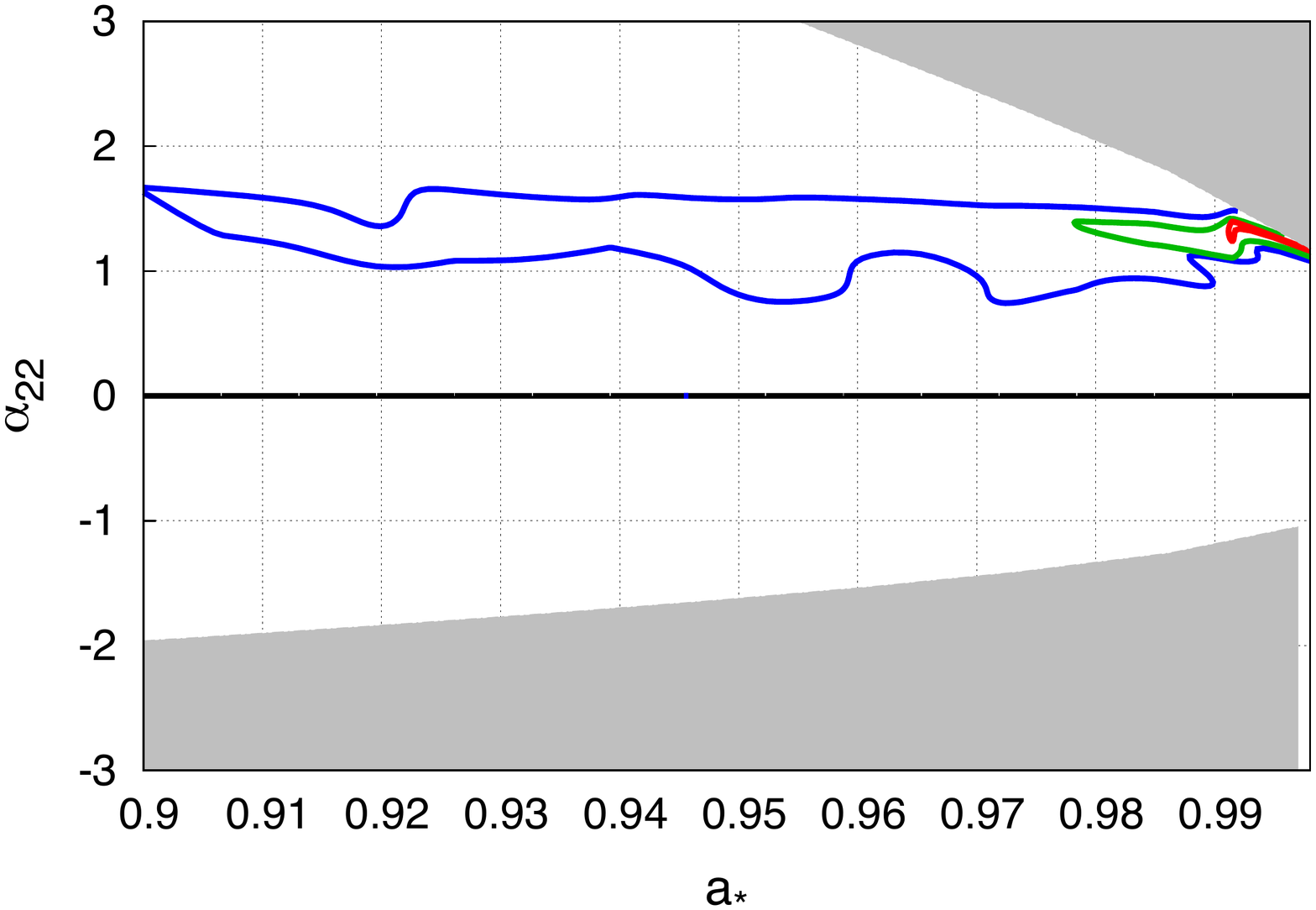}
\end{center}
\vspace{-0.6cm}
\caption{Constraints on the spin parameter $a_*$ and the Johannsen deformation parameters $\alpha_{13}$ (left panel) and $\alpha_{22}$ (right panel) from epoch~4. The emissivity profile is modeled with a power-law (top panels), a broken power-law with outer emissivity index frozen to 3 (central panels), and a broken power-law with both emissivity indices free (bottom panels). The red, green, and blue curves are, respectively, the 68\%, 90\%, and 99\% confidence level boundaries for two relevant parameters. \label{f-c4}}
\end{figure*}

\begin{figure*}[t]
  \begin{center}
  \includegraphics[width=8.5cm,trim={0.5cm 1.0cm 0.5cm 1.0cm},clip]{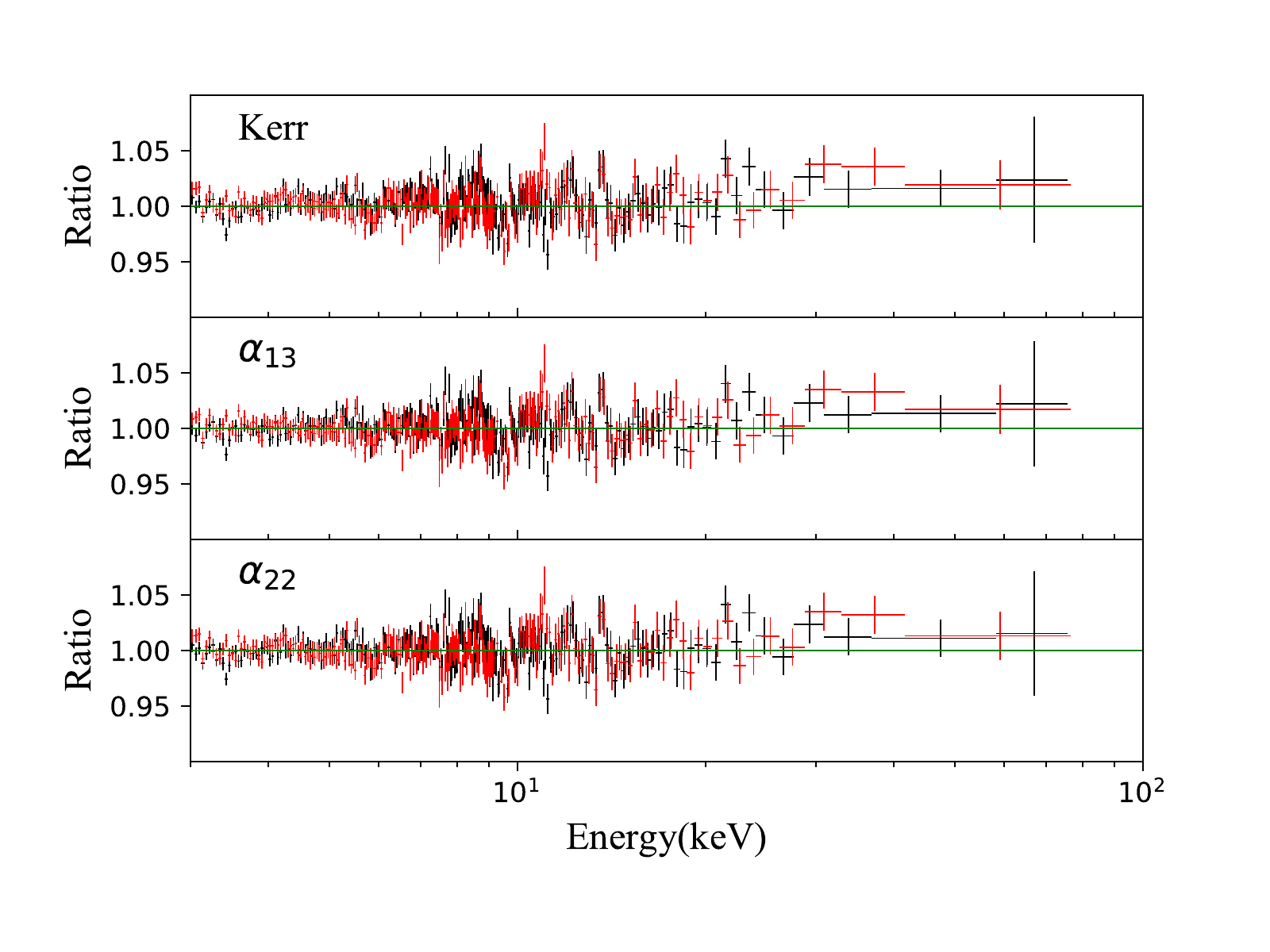}
  \includegraphics[width=8.5cm,trim={0.5cm 1.0cm 0.5cm 1.0cm},clip]{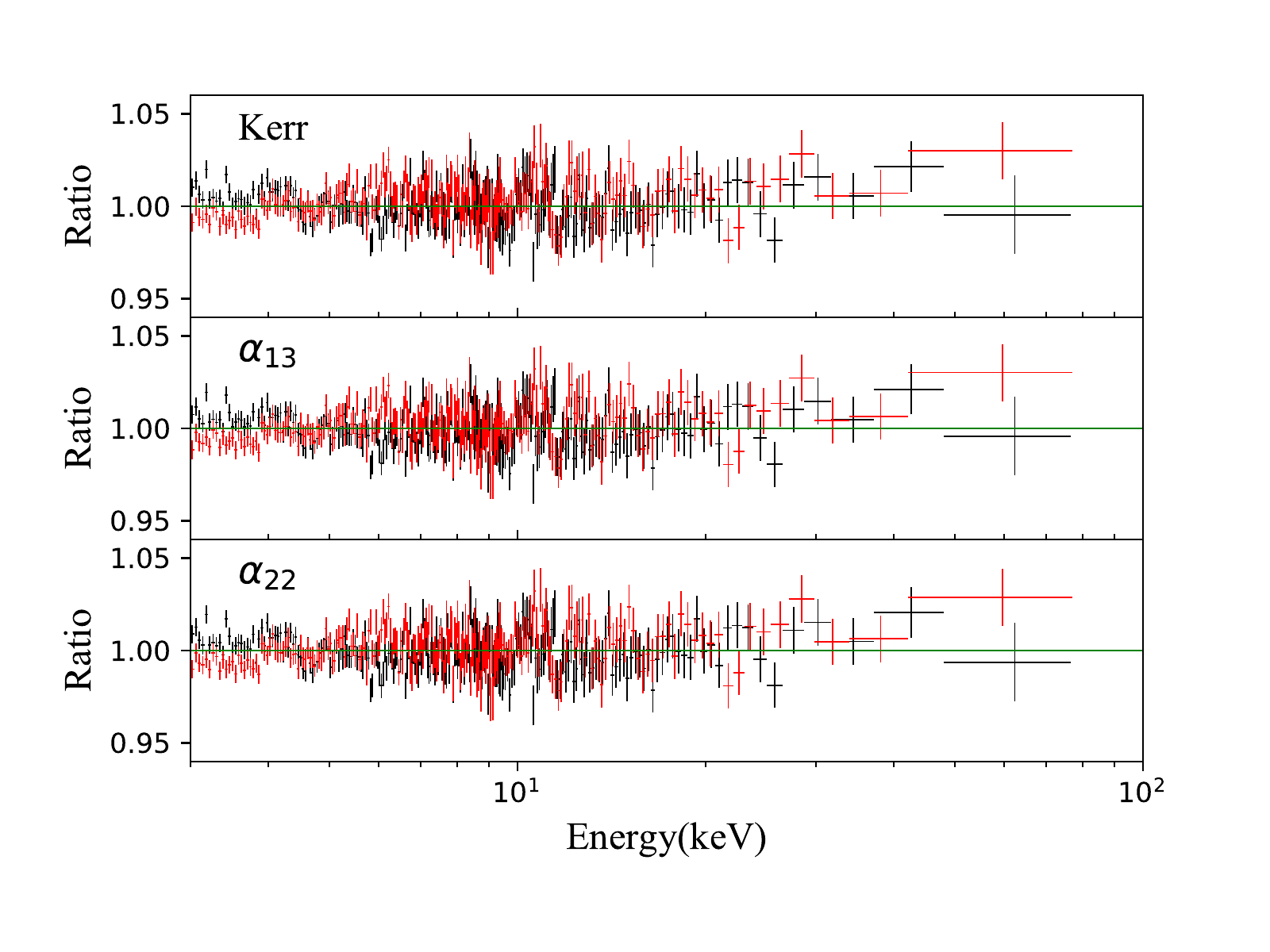}
  \end{center}
  \vspace{-0.4cm}
  \caption{Ratio plots for epoch~1 (left panel) and epoch~4 (right panel). The emissivity profile is modeled with a power-law. \label{f-ratio-a}}
\end{figure*}

\begin{figure*}[t]
  \begin{center}
  \includegraphics[width=8.5cm,trim={1.0cm 1.0cm 2.0cm 1.0cm},clip]{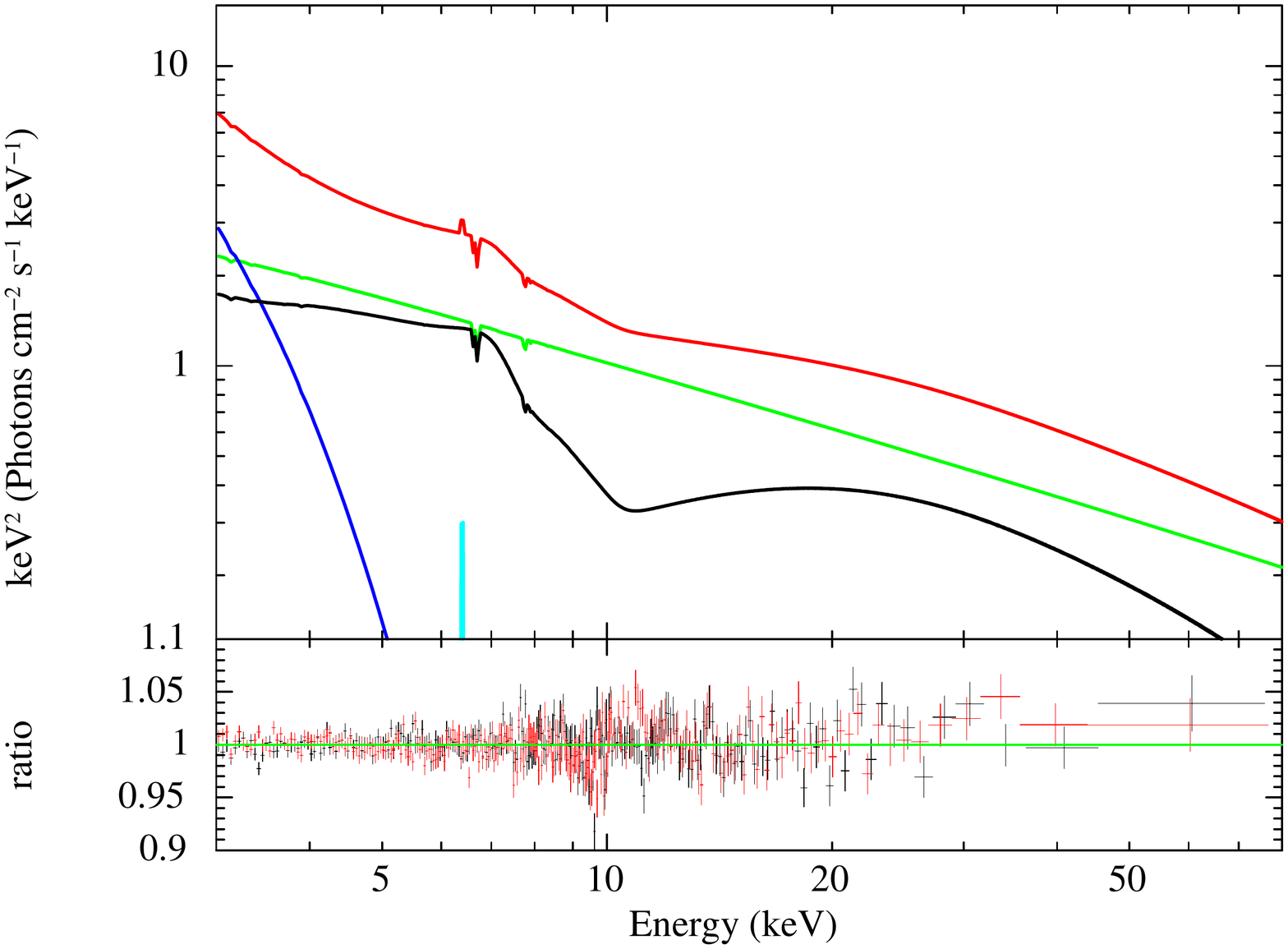}
  \includegraphics[width=8.5cm,trim={1.0cm 1.0cm 2.0cm 1.0cm},clip]{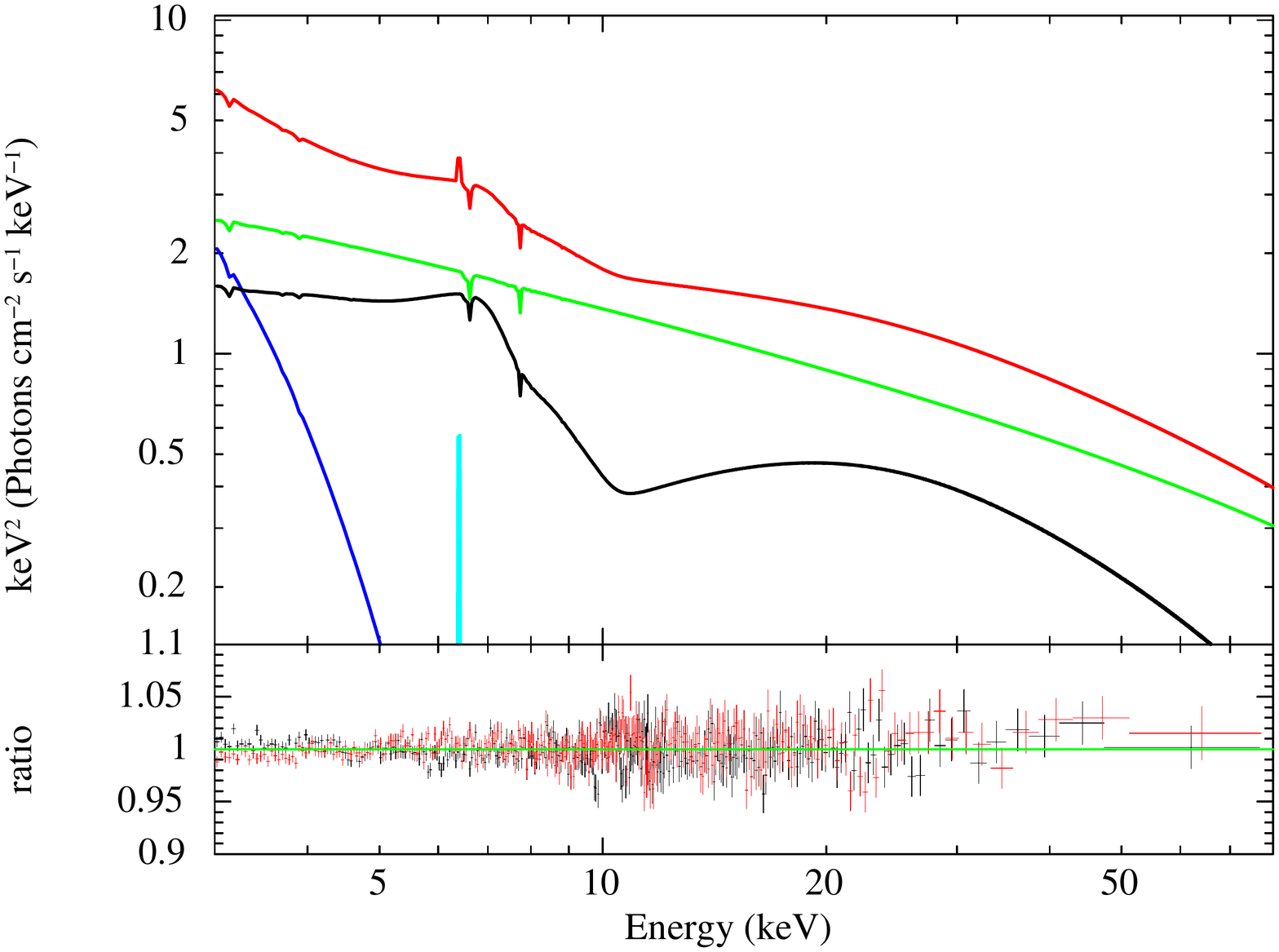} \\ \vspace{0.4cm}
  \includegraphics[width=8.5cm,trim={1.0cm 1.0cm 2.0cm 1.0cm},clip]{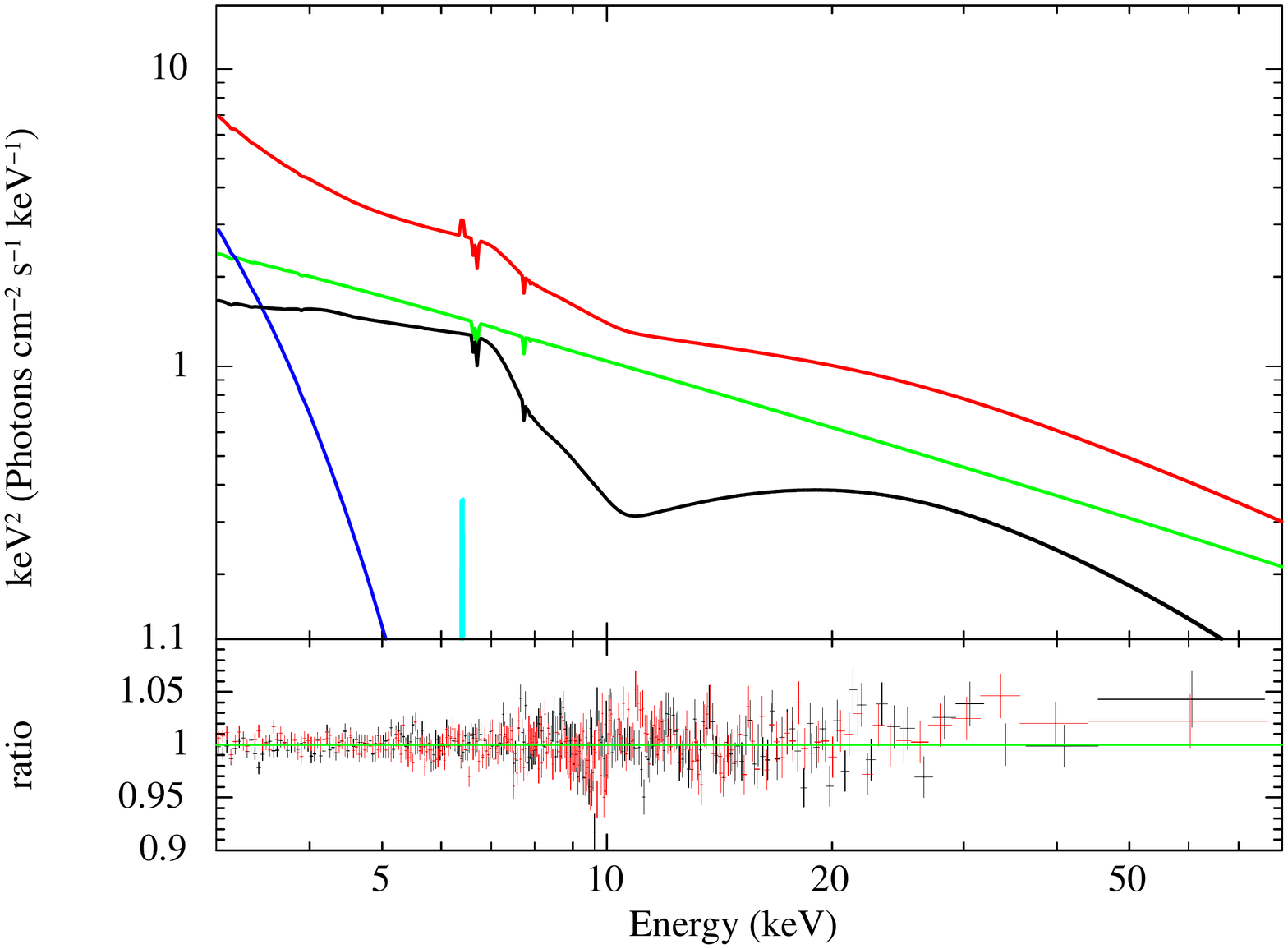}
  \includegraphics[width=8.5cm,trim={1.0cm 1.0cm 2.0cm 1.0cm},clip]{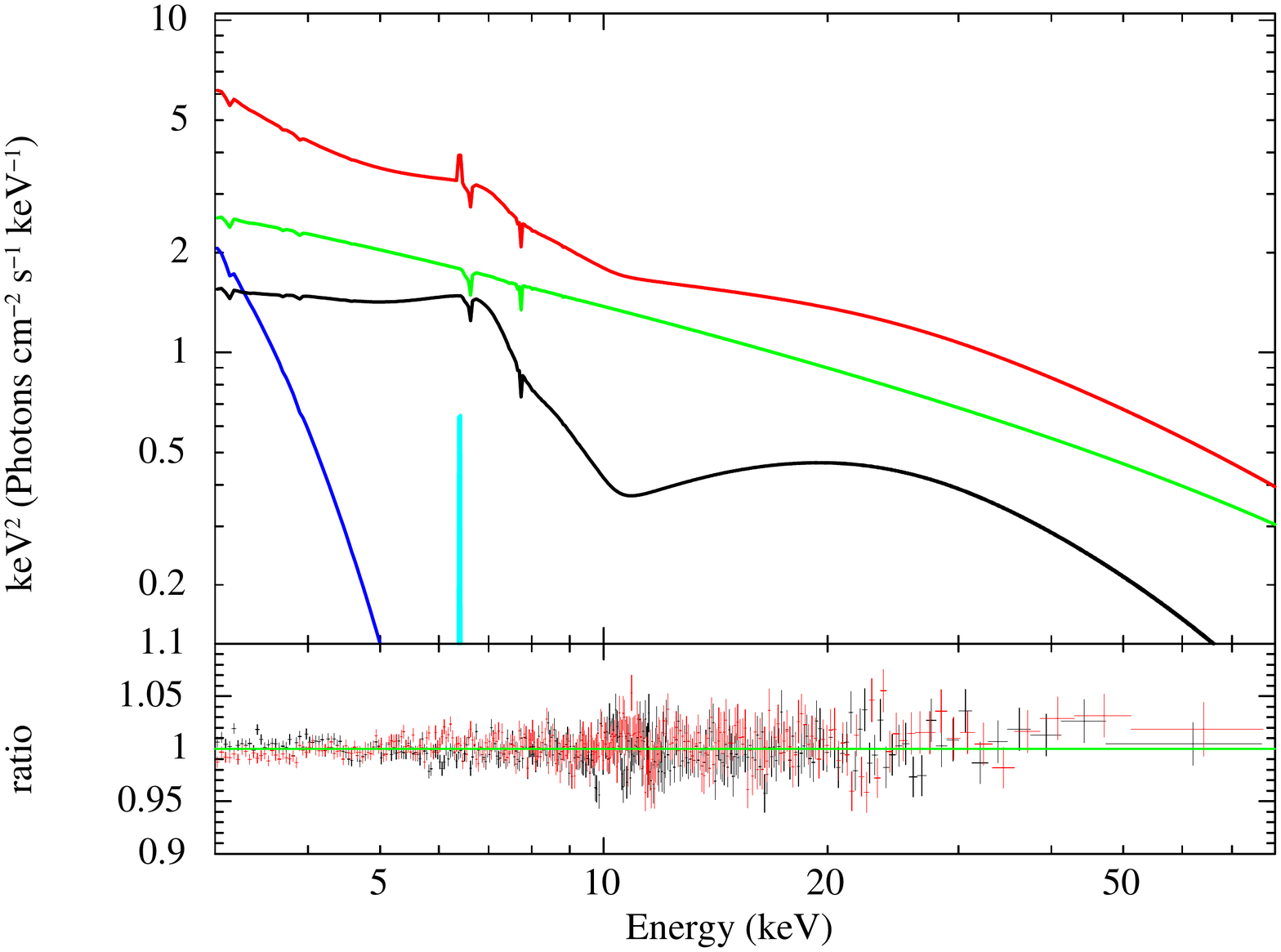} \\ \vspace{0.4cm}
  \includegraphics[width=8.5cm,trim={1.0cm 1.0cm 2.0cm 1.0cm},clip]{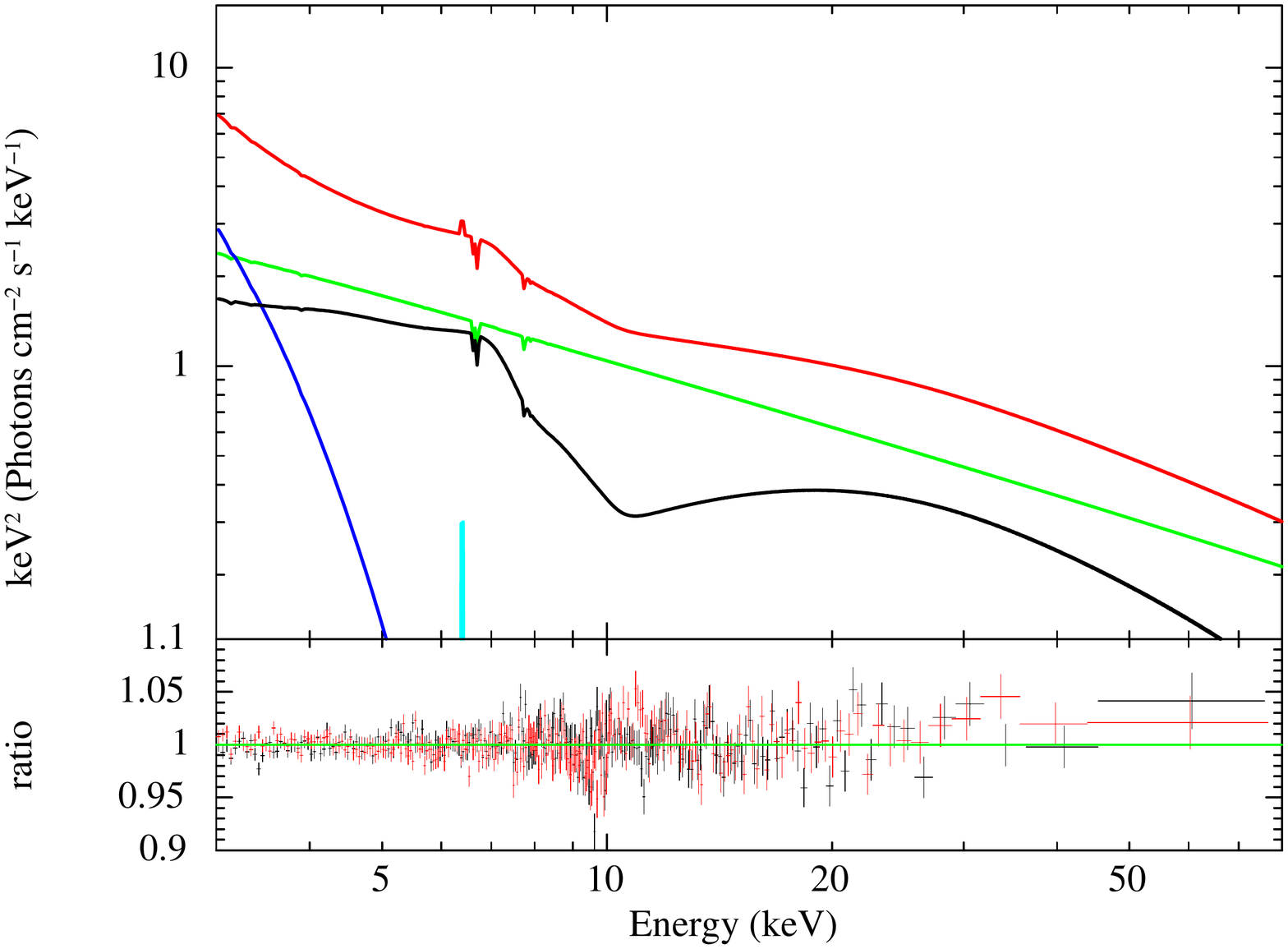}
  \includegraphics[width=8.5cm,trim={1.0cm 1.0cm 2.0cm 1.0cm},clip]{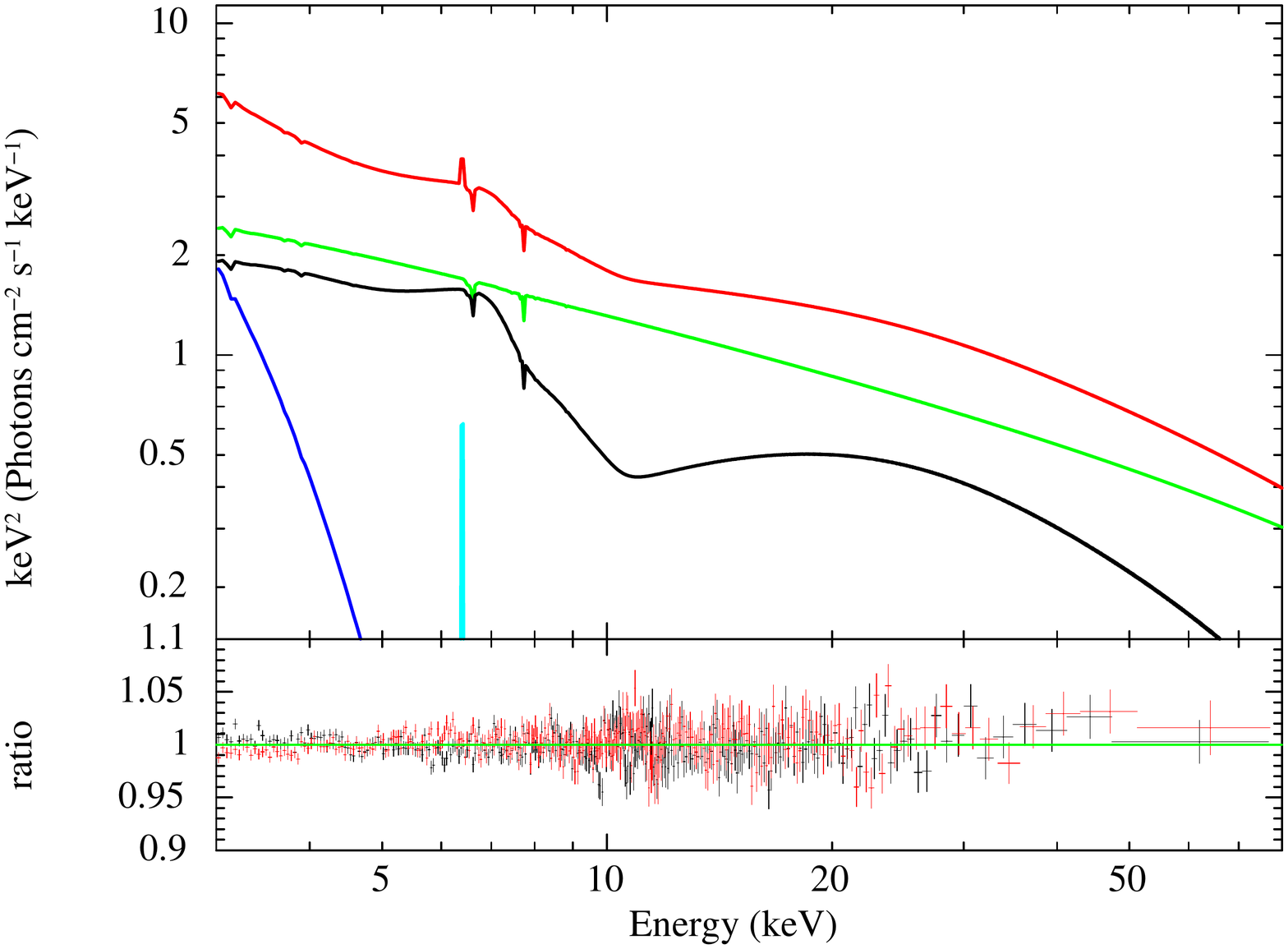}
  \end{center}
  \vspace{-0.4cm}
  \caption{Best-fit models and ratio plots for epoch~1 (left panel) and epoch~4 (right panel), assuming $\alpha_{13} = \alpha_{22} = 0$ (Kerr metric, top panels), $\alpha_{13}$ free and $\alpha_{22} = 0$ (central panels), and $\alpha_{13} = 0$ and $\alpha_{22}$ free (bottom panels). The emissivity profile is modeled with a broken power-law with outer emissivity index frozen to 3. \label{f-ratio-b}}
\end{figure*}

\begin{figure*}[t]
  \begin{center}
  \includegraphics[width=8.5cm,trim={0.5cm 1.0cm 0.5cm 1.0cm},clip]{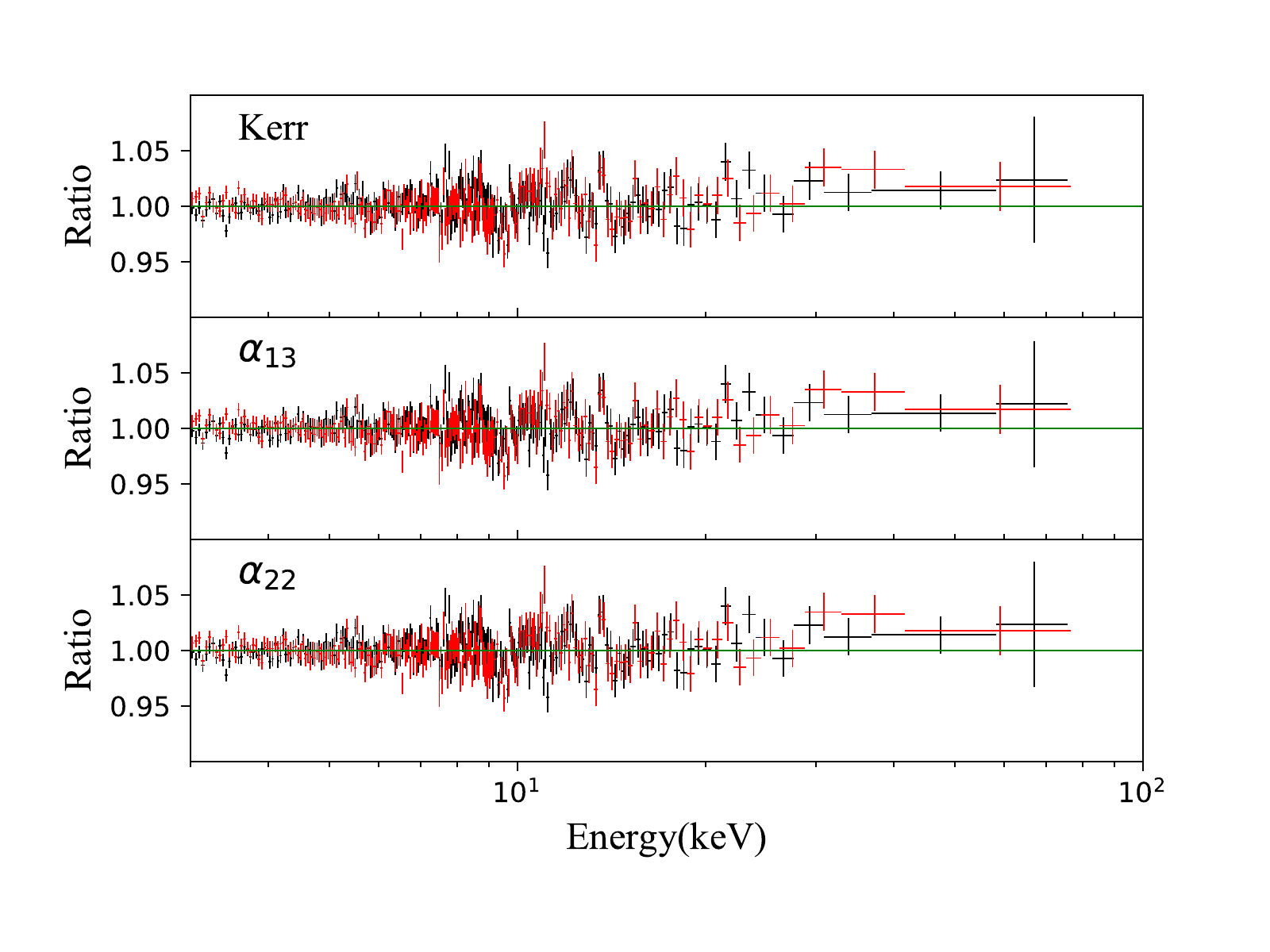}
  \includegraphics[width=8.5cm,trim={0.5cm 1.0cm 0.5cm 1.0cm},clip]{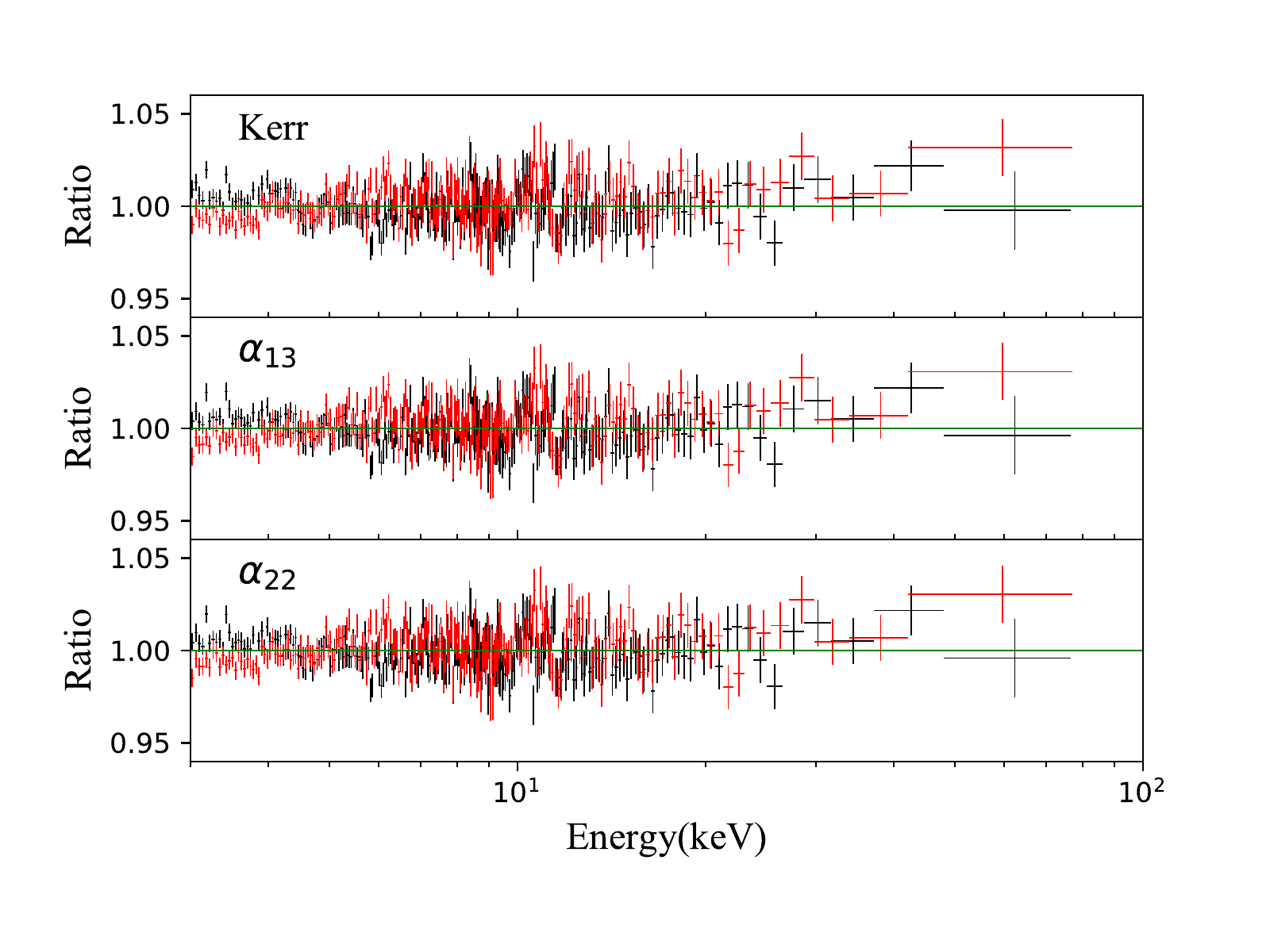}
  \end{center}
  \vspace{-0.4cm}
  \caption{Ratio plots for epoch~1 (left panel) and epoch~4 (right panel). The emissivity profile is modeled with a broken power-law with both emissivity indices free. \label{f-ratio-c}}
\end{figure*}

Here we report the results of our analysis of the observations of epoch~1 and 4 (we have analyzed the data of epoch~2 and 3 as well, but the quality of the data is clearly less suitable for our tests and we have not performed a detailed analysis of them). For every epoch, we consider three models for the intensity profile: power-law, broken power-law with $q_{\rm out} = 3$, broken power-law with $q_{\rm out}$ free. For every model of the intensity profile, we consider three spacetimes: Kerr metric, Johannsen metric with $\alpha_{13}$ free and $\alpha_{22} = 0$, and Johannsen metric with $\alpha_{13} = 0$ and $\alpha_{22}$ free. We thus have 18~fits in total. The best-fit values for the model with a power-law profile are shown in Tab.~\ref{t-fit1}. Those for the models with a broken power-law and $q_{\rm out} = 3$ are shown in Tab.~\ref{t-fit2}, and those for the models with a broken power-law and $q_{\rm out}$ free are shown in Tab.~\ref{t-fit3}. The constraints on the spin parameter and the deformation parameters from the 12~non-Kerr fits are shown in Figs.~\ref{f-c1} and \ref{f-c4}, respectively for the epoch~1 and 4. The plots of the data to best-fit model ratios for the 6~fits with a simple power-law are shown in Fig.~\ref{f-ratio-a}, the best-fit models and the ratio plots for the 6~fits with a broken power-law and $q_{\rm out} = 3$ are shown in Fig.~\ref{f-ratio-b}, and the plots of the data to best-fit model ratios for the 6~fits with a broken power-law and $q_{\rm out}$ free are shown in Fig.~\ref{f-ratio-c}.

\begin{figure*}[t]
  \begin{center}
  \includegraphics[width=16cm]{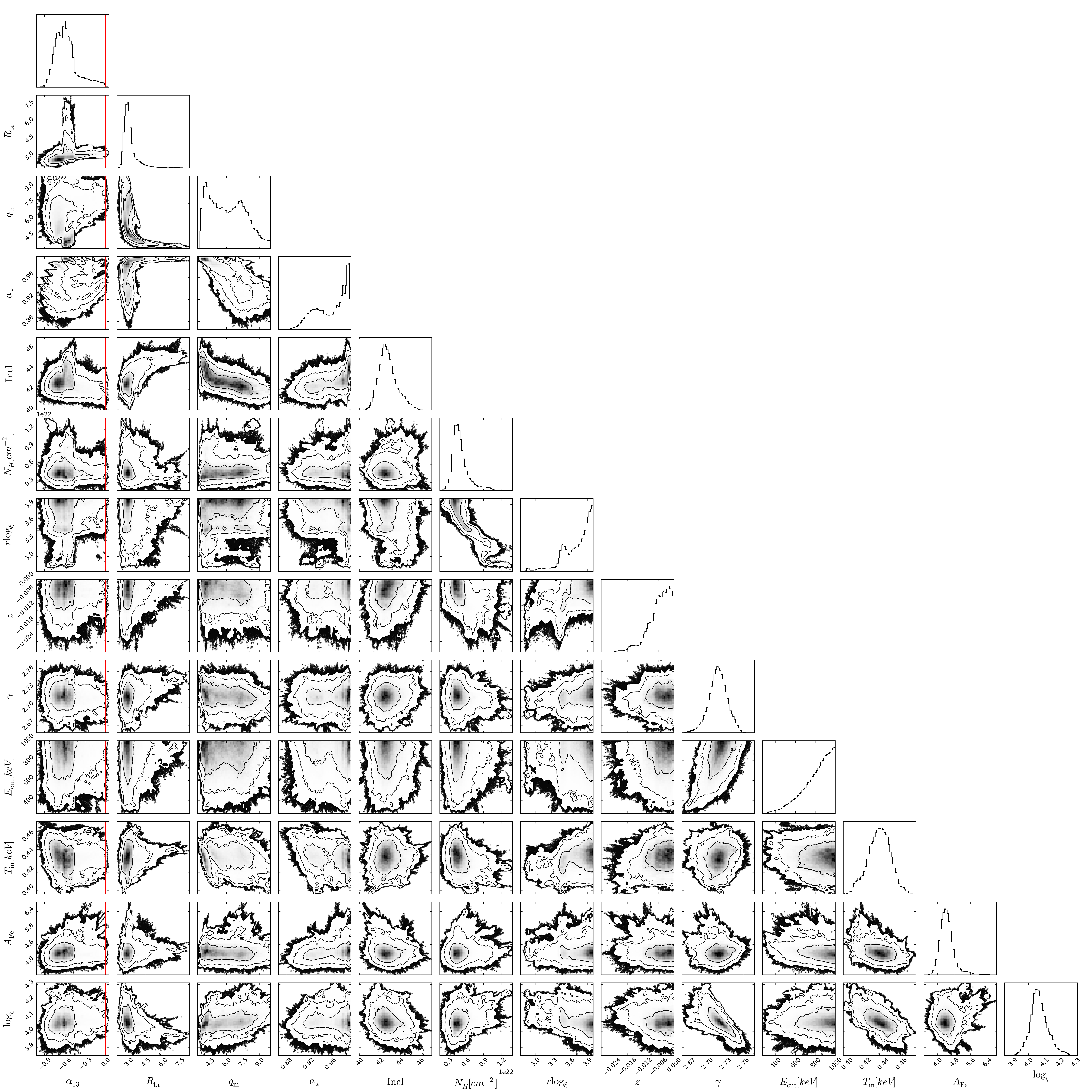}
  \end{center}
  \vspace{0.0cm}
  \caption{Output distribution of the MCMC analysis for epoch~1. The contours correspond to 1-, 2-, and 3-$\sigma$ confidence level curves. Note that here $N_{\rm H}$ and $z$ refer, respectively, to the column density and the observed redshift of the absorber in {\sc xstar} ($1 + z = \sqrt{(1-\beta)/(1+\beta)}$, where $\beta=v_{\rm out}/c$). } \label{mcmc-e1}
\end{figure*}

\begin{figure*}[t]
  \begin{center}
  \includegraphics[width=16cm]{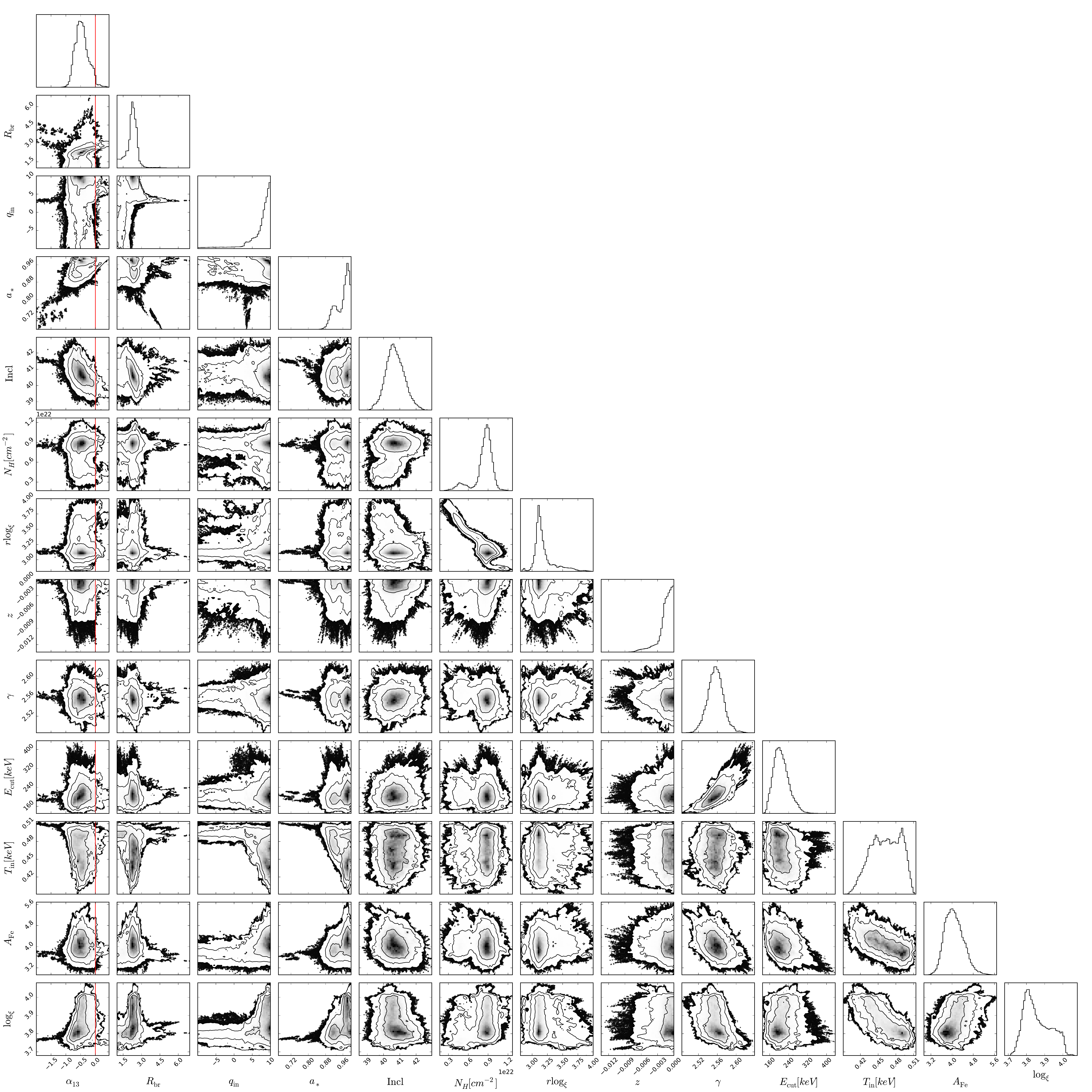}
  \end{center}
  \vspace{0.0cm}
  \caption{Output distribution of the MCMC analysis for epoch~4. The contours correspond to 1-, 2-, and 3-$\sigma$ confidence level curves. Note that here $N_{\rm H}$ and $z$ refer, respectively, to the column density and the observed redshift of the absorber in {\sc xstar} ($1 + z = \sqrt{(1-\beta)/(1+\beta)}$, where $\beta=v_{\rm out}/c$).} \label{mcmc-e4}
\end{figure*}

\begin{figure*}[t]
  \begin{center}
  \includegraphics[width=8.5cm,clip]{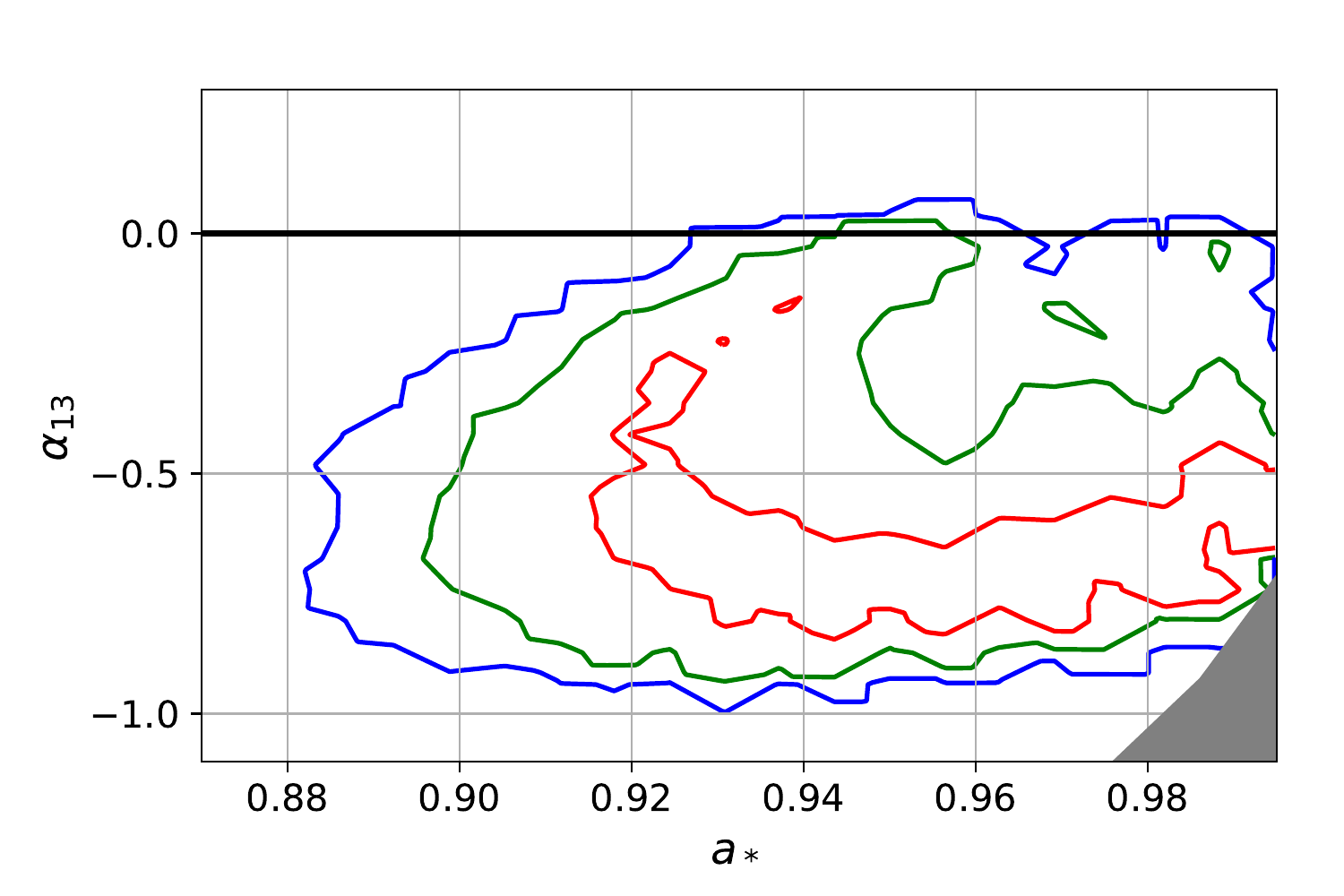}
  \includegraphics[width=8.5cm,clip]{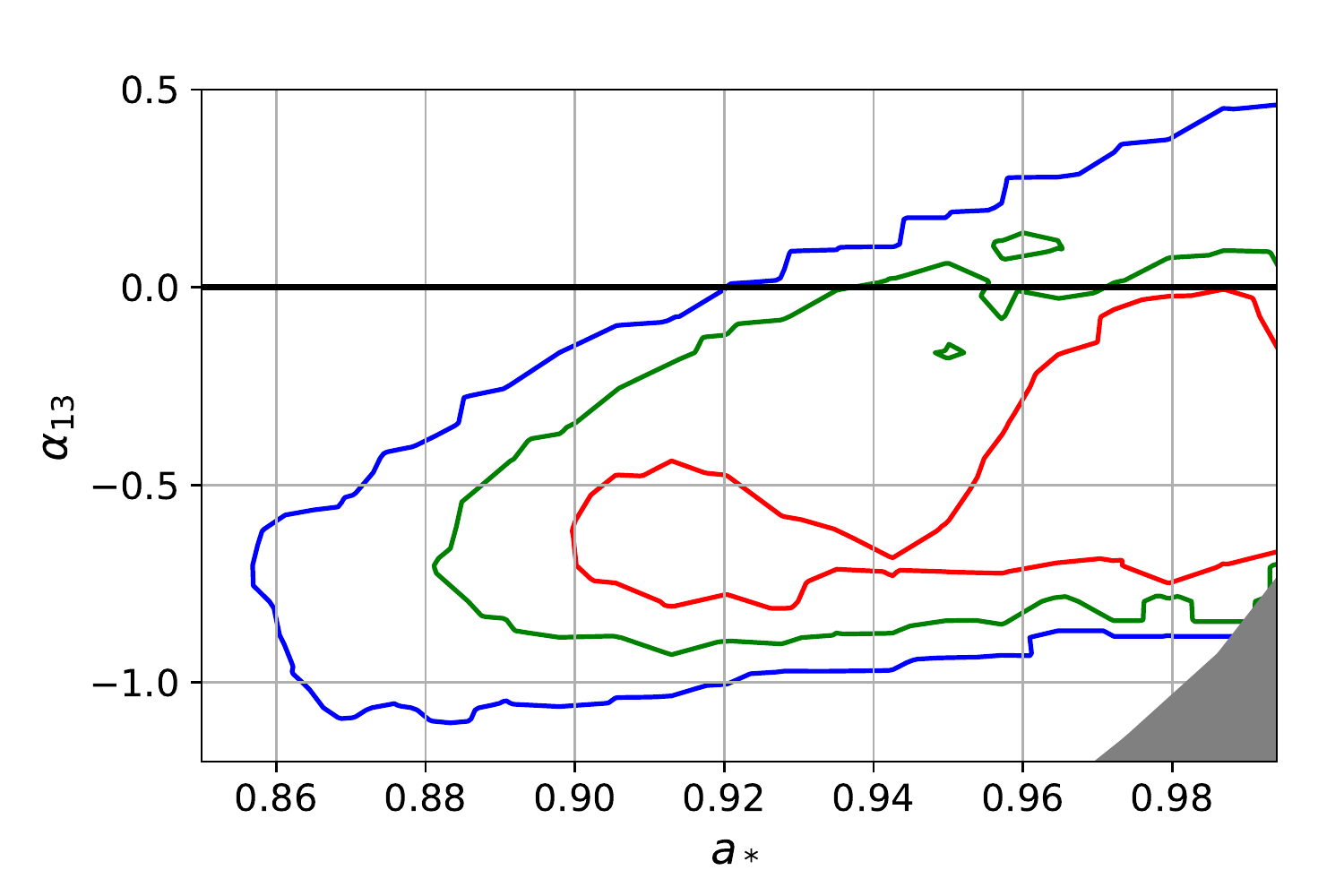}
  \end{center}
  \vspace{0.0cm}
  \caption{Constraints on the spin parameter $a_*$ and the deformation parameter $\alpha_{13}$ from the analysis with MCMC simulations for the spectra of epoch~1 (left panel) and epoch~4 (right panel). The red, green, and blue lines correspond to 1-, 2-, and 3-$\sigma$ confidence level curves, respectively. \label{f-ccc}}
\end{figure*}

Before discussing the results of these fits in the next section, we show a more detailed analysis of the two cases in which we assume a broken power-law with $q_{\rm out} = 3$ and the metric is described by the Johannsen solution with $\alpha_{13}$ free (one is for epoch~1, the other is for epoch~4). For these two fits, we run MCMC simulations with the ``chain'' command in XSPEC. We use 100~walkers with 1.5~million steps each (about 15~times the autocorrelation length), burning the first 0.5~million. We also compared the distributions from the first half and the second half of the samples and we do not find large differences. The results of these simulations are shown in Fig.~\ref{mcmc-e1} and Fig.~\ref{mcmc-e4}, respectively for epoch~1 and 4. Fig.~\ref{f-ccc} shows the constraints on the spin parameter $a_*$ and the deformation parameter $\alpha_{13}$ from this analysis with MCMC simulations.


\begin{table*}
\centering
\vspace{0.5cm}
\begin{tabular}{lccc|ccc}
\hline\hline
& \multicolumn{3}{c}{Epoch~1} & \multicolumn{3}{c}{Epoch~4} \\
Model & Kerr & $\alpha_{13}$ & $\alpha_{22}$ & Kerr & $\alpha_{13}$ & $\alpha_{22}$ \\
\hline
{\sc tbabs} && \\
$N_{\rm H} / 10^{21}$ cm$^{-2}$ & $6.0^\star$ & $6.0^\star$ & $6.0^\star$ & $6.0^\star$ & $6.0^\star$ & $6.0^\star$ \\
\hline
{\sc xstar} && \\
$N_{\rm H} / 10^{21}$ cm$^{-2}$ & $4.0^{+1.2}_{-1.0}$ & $4.1_{-1.4}^{+0.4}$ & $3.5^{+0.5}_{-1.0}$ & $9.4^{+0.9}_{-1.5}$ & $9.3^{+1.1}_{-1.2}$ & $9.4^{+1.1}_{-1.5}$ \\
$\log\xi$ & $3.74_{-0.18}^{+0.15}$ &$3.76_{-0.06}^{+0.08} $& $3.90^{+0.07}_{-0.22}$ & $3.11^{+0.07}_{-0.06}$ & $3.08^{+0.11}_{-0.06}$ & $3.10^{+0.10}_{-0.06}$ \\
$v_{\rm out} /$ km s$^{-1}$ &$<1800$&$<1200 $&$<1500 $&$<600 $&$<600$& $<900$ \\
\hline
{\sc cutoffpl} && \\
$\Gamma$ & $2.74^{+0.02}_{-0.02}$ & $2.741^{+0.015}_{-0.020}$ & $2.736^{+0.018}_{-0.018}$ & $2.550^{+0.019}_{-0.016}$ & $2.550^{+0.016}_{-0.007}$ & $2.551^{+0.017}_{-0.026}$ \\
$E_{\rm cut}$ [keV] & $>746$ & $>903$ & $>900$ & $184^{+28}_{-22}$ & $185^{+6}_{-4}$ & $188^{+26}_{-35}$ \\
{Norm} & $5.9_{-0.3}^{+0.4}$ &$5.9_{-0.3}^{+0.3} $&$5.862_{-0.021}^{+0.4} $&$5.23_{-0.12}^{+0.24} $&$5.23_{-0.09}^{+0.20} $& $5.22_{-0.13}^{+0.14}$ \\
\hline
{\sc diskbb} && \\
$T_{\rm in}$ [keV] & $0.476^{+0.007}_{-0.007}$ & $0.448^{+0.005}_{-0.007}$ & $0.473^{+0.006}_{-0.003}$ & $0.498^{+0.008}_{-0.005}$ & $0.478^{+0.006}_{-0.002}$ & $0.491^{+0.007}_{-0.006}$ \\
Norm $[10^4]$ & $3.80_{-0.4}^{+0.17} $ &$5.9_{-0.6}^{+0.9} $ &$4.0_{-0.4}^{+0.5} $& $ 1.97_{-0.22}^{+0.18} $ & $ 2.66_{-0.06}^{+0.016} $ &$ 2.19_{-0.3}^{+0.18} $ \\
\hline
{\sc relconv\_nk} && \\
$q_{\rm in}$ & $4.15^{+0.19}_{-0.21}$ & $3.41_{-0.12}^{+0.12} $ & $3.251^{+0.014}_{-0.08}$ & $3.15^{+0.09}_{-0.08}$ & $2.889_{-0.05}^{+0.009}$ & $2.87^{+0.10}_{-0.06}$ \\
$q_{\rm out}$ & $= q_{\rm in}$ & $= q_{\rm in}$ & $= q_{\rm in}$ & $= q_{\rm in}$ & $= q_{\rm in}$ & $= q_{\rm in}$\\
$R_{\rm br}$~$[M]$ & -- & -- & -- & -- & -- & -- \\
$a_*$ & $ 0.967_{-0.010}^{+0.005} $ & $0.9954_{-0.0009}^{+0.0007}$ & $>0.997$ & $>0.93$ & $0.9918_{-0.0003}^{+0.0011}$ & $> 0.990$ \\
$i$ [deg] & $47.2^{+1.3}_{-1.6}$ & $45.3^{+0.3}_{-0.9}$ & $44.3^{+0.4}_{-0.4}$ & $41.0^{+0.5}_{-0.9}$ & $40.8^{+0.6}_{-0.3}$ & $40.7^{+0.6}_{-0.6}$ \\
$\alpha_{13}$ & $0^\star$ & $<-0.79 $ & $0^\star$ & $0^\star$ & $<-0.87$ & $0^\star$ \\
$\alpha_{22}$ & $0^\star$ & $0^\star$ & $>0.85 $ & $0^\star$ & $0^\star$ & $>0.59$ \\
\hline
{\sc xillver} && \\
$A_{\rm Fe}$ & $4.1^{+0.4}_{-0.4}$ & $4.30^{+0.38}_{-0.05}$ & $4.01^{+0.05}_{-0.05}$ & $3.79^{+0.3}_{-0.23}$ & $3.96^{+0.05}_{-0.21}$ & $3.80_{-0.27}^{+0.27}$ \\
$\log\xi$ & $3.93^{+0.09}_{-0.05}$ & $4.000^{+0.005}_{-0.05}$ & $3.970^{+0.015}_{-0.014}$ & $3.80^{+0.03}_{-0.04}$ & $3.817^{+0.010}_{-0.027}$ & $3.81^{+0.04}_{-0.03}$ \\
{Norm} & $0.127_{-0.010}^{+0.009} $ &$0.133_{-0.012}^{+0.022} $&$0.1159_{-0.012}^{+0.0007} $& $ 0.064_{-0.004}^{+0.003} $ & $0.067_{-0.005}^{+0.002} $& $ 0.065_{-0.003}^{+0.003} $ \\
\hline
{\sc gaussian} && \\
{Norm} $[10^{-4}]$ &$6.9_{-1.0}^{+1.7} $ &$7.7_{-2.2}^{+1.3} $&$ 6.7_{-1.3}^{+1.3} $& $ 12_{-2}^{+2} $ &$12.5_{-1.2}^{+1.8} $& $ 12_{-2}^{+2} $ \\
\hline
$\chi^2/\nu$ & $1764.82/1605 \quad$ & $\quad 1721.06/1604 \quad$ & $\quad 1742.10/1604 \quad$ & $\quad 2092.82/1913 \quad$ & $\quad 2080.72/1912 \quad$ & $\quad 2091.77/1912 \quad$ \\
& =1.09957 & =1.07298 & =1.0861 & =1.094 & 1.08824 & 1.09402 \\
\hline\hline
\end{tabular}
\vspace{0.2cm}
\caption{Summary of the best-fit values assuming a power-law emissivity profile. The reported uncertainties correspond to the 90\% confidence level for one relevant parameter. $^\star$ indicates that the parameter is frozen. \label{t-fit1}}
\end{table*}


\begin{table*}
  \centering
  \vspace{0.5cm}
  \begin{tabular}{lccc|ccc}
  \hline\hline
  & \multicolumn{3}{c}{Epoch~1} & \multicolumn{3}{c}{Epoch~4} \\
  Model & Kerr & $\alpha_{13}$ & $\alpha_{22}$ & Kerr & $\alpha_{13}$ & $\alpha_{22}$ \\
  \hline
  {\sc tbabs} && \\
 $N_{\rm H} / 10^{21}$ cm$^{-2}$ & $6.0^\star$ & $6.0^\star$ & $6.0^\star$ & $6.0^\star$ & $6.0^\star$ & $6.0^\star$ \\
  \hline
  {\sc xstar} && \\
  $N_{\rm H} / 10^{21}$ cm$^{-2}$ & $4.7^{+0.9}_{-1.2}$ & $4.1_{-1.1}^{+0.9}$ & $4.4^{+0.6}_{-0.8}$ & $9.6^{+1.3}_{-1.3}$ & $9.2^{+1.1}_{-1.3}$ & $9.2^{+1.1}_{-1.4}$ \\
  $\log\xi$ & $>3.75$ &$>3.50$& $>3.81$ & $3.10^{+0.14}_{-0.11}$ & $3.09^{+0.15}_{-0.08}$ & $3.10^{+0.10}_{-0.08}$ \\
  $v_{\rm out} /$ km s$^{-1}$ &$<2700$&$<2100 $&$<1800 $&$<600 $&$<600 $& $<600$ \\
  \hline
  {\sc cutoffpl} && \\
  $\Gamma$ & $2.725^{+0.019}_{-0.023}$ & $2.735^{+0.008}_{-0.020}$ & $2.733^{+0.001}_{-0.008}$ & $2.540^{+0.018}_{-0.03}$ & $2.548^{+0.024}_{-0.012}$ & $2.546^{+0.025}_{-0.016}$ \\
  $E_{\rm cut}$ [keV] & $>593$ & $>840$ & $>905$ & $184^{+32}_{-15}$ & $184^{+21}_{-22}$ & $206^{+14}_{-13}$ \\
  {Norm} & $5.5_{-0.4}^{+0.4}$ &$5.7_{-0.5}^{+0.3} $&$5.71_{-0.021}^{+0.05} $&$5.01_{-0.21}^{+0.17} $&$5.18_{-0.13}^{+0.3} $& $4.89_{-0.27}^{+0.3}$ \\
  \hline
  {\sc diskbb} && \\
  $T_{\rm in}$ [keV] & $0.441^{+0.015}_{-0.018}$ & $0.437^{+0.006}_{-0.008}$ & $0.439^{+0.024}_{-0.005}$ & $0.470^{+0.013}_{-0.022}$ & $0.469^{+0.03}_{-0.026}$ & $0.430^{+0.019}_{-0.028}$ \\
  Norm $[10^4]$ & $6.7_{-1.5}^{+3.0} $ &$7.3_{-1.7}^{+2.6} $ &$7.03_{-0.24}^{+0.3} $& $ 3.0_{-0.6}^{+1.1} $ & $ 3.0_{-0.8}^{+0.4} $ &$ 5.4_{-0.7}^{+3.0} $ \\
  \hline
  {\sc relconv\_nk} && \\
  $q_{\rm in}$ & $7.8^{+1.3}_{-1.0}$ & $4.3_{-0.4}^{+1.8} $ & $5.05^{+0.04}_{-0.4}$ & $>7.9$ & $>8.8$ & $>7.8$ \\
  $q_{\rm out}$ & $3^\star$ & $3^\star$ & $3^\star$ &$3^\star$ & $3^\star$ & $3^\star$\\
  $R_{\rm br}$~$[M]$ & $3.25_{-0.22}^{+0.26} $ & $ 3.3_{-0.7}^{+1.0} $ & $2.596_{-0.024}^{+0.19} $ & $ 2.63_{-0.06}^{+0.14} $ & $ 2.19_{-0.23}^{+0.27} $ & $ 2.02_{-0.15}^{+0.24} $ \\
  $a_*$ & $ 0.951_{-0.012}^{+0.009} $ & $0.989_{-0.014}^{+0.003}$ & $>0.997$ & $0.945_{-0.009}^{+0.022}$ & $0.924_{-0.03}^{+0.027}$ & $0.988_{-0.012}^{+0.005}$ \\
  $i$ [deg] & $41.8^{+0.8}_{-0.7}$ & $43.3^{+1.1}_{-1.0}$ & $42.7^{+0.3}_{-0.3}$ & $39.8^{+0.5}_{-0.9}$ & $41.0^{+0.4}_{-0.4}$ & $41.0^{+0.4}_{-0.5}$ \\
  $\alpha_{13}$ & $0^\star$ & $-0.79_{-0.08}^{+0.09} $ & $0^\star$ & $0^\star$ & $-0.40_{-0.23}^{+0.3}$ & $0^\star$ \\
  $\alpha_{22}$ & $0^\star$ & $0^\star$ & $0.95_{-0.05}^{+0.04} $ & $0^\star$ & $0^\star$ & $0.5_{-0.5}^{+0.4}$ \\
  \hline
  {\sc xillver} && \\
  $A_{\rm Fe}$ & $4.2^{+0.4}_{-0.4}$ & $4.31^{+0.21}_{-0.4}$ & $4.31^{+0.05}_{-0.4}$ & $3.9^{+0.3}_{-0.3}$ & $4.04^{+0.4}_{-0.27}$ & $4.5_{-0.4}^{+0.5}$ \\
  $\log\xi$ & $4.06^{+0.08}_{-0.05}$ & $4.030^{+0.012}_{-0.013}$ & $4.037^{+0.005}_{-0.05}$ & $3.88^{+0.05}_{-0.03}$ & $3.828^{+0.027}_{-0.04}$ & $3.98^{+0.04}_{-0.03}$ \\
  {Norm} & $0.115_{-0.013}^{+0.016} $ &$0.126_{-0.008}^{+0.021} $&$0.123_{-0.010}^{+0.0008} $& $ 0.066_{-0.006}^{+0.005} $ & $0.069_{-0.008}^{+0.010} $& $ 0.092_{-0.004}^{+0.029} $ \\
  \hline
  {\sc gaussian} && \\
  {Norm} $[10^{-4}]$ &$5.9_{-3.0}^{+2.8} $ &$7.0_{-1.3}^{+2.1} $&$ 5.9_{-1.3}^{+1.3} $& $ 12_{-2}^{+2} $ &$12.7_{-2.4}^{+1.8} $& $ 12.2_{-1.5}^{+1.9} $ \\
  \hline
  $\chi^2/\nu$ & $1721.47/1604 \quad$ & $\quad 1716.64/1603 \quad$ & $\quad 1720.79/1603 \quad$ & $\quad 2085.54/1912 \quad$ & $\quad 2082.81/1911 \quad$ & $\quad 2082.34/1911 \quad$ \\
  & =1.07324 & =1.07089 & =1.07348 & =1.09076 & 1.08991 & 1.08966 \\
  \hline\hline
  \end{tabular}
  \vspace{0.2cm}
  \caption{Summary of the best-fit values assuming a broken power-law emissivity profile with $q_{\rm out} = 3$. The reported uncertainties correspond to the 90\% confidence level for one relevant parameter. $^\star$ indicates that the parameter is frozen. \label{t-fit2}}
  \end{table*}


\begin{table*}
  \centering
  \vspace{0.5cm}
  \begin{tabular}{lccc|ccc}
  \hline\hline
  & \multicolumn{3}{c}{Epoch~1} & \multicolumn{3}{c}{Epoch~4} \\
  Model & Kerr & $\alpha_{13}$ & $\alpha_{22}$ & Kerr & $\alpha_{13}$ & $\alpha_{22}$ \\
  \hline
  {\sc tbabs} && \\
  $N_{\rm H} / 10^{21}$ cm$^{-2}$ & $6.0^\star$ & $6.0^\star$ & $6.0^\star$ & $6.0^\star$ & $6.0^\star$ & $6.0^\star$ \\
  \hline
  {\sc xstar} && \\
  $N_{\rm H} / 10^{21}$ cm$^{-2}$ & $4.2^{+1.0}_{-1.3}$ & $4.2_{-1.1}^{+0.9}$ & $4.3^{+0.7}_{-1.3}$ & $9.3^{+1.2}_{-1.7}$ & $9.2^{+1.4}_{-0.9}$ & $9.3^{+1.3}_{-1.8}$ \\
  $\log\xi$ & $>3.46$ &$>3.47$& $>3.51$ & $3.09^{+0.24}_{-0.10}$ & $3.06^{+0.25}_{-0.10}$ & $3.08^{+0.27}_{-0.11}$ \\
  $v_{\rm out} /$ km s$^{-1}$ &$<2100$&$<2100 $&$<2100 $&$<900 $&$<900 $& $<900$ \\
  \hline
  {\sc cutoffpl} && \\
  $\Gamma$ & $2.738^{+0.018}_{-0.024}$ & $2.736^{+0.009}_{-0.04}$ & $2.738^{+0.014}_{-0.026}$ & $2.548^{+0.026}_{-0.027}$ & $2.546^{+0.011}_{-0.016}$ & $2.541^{+0.023}_{-0.029}$ \\
  $E_{\rm cut}$ [keV] & $>722$ & $>729$ & $>708$ & $187^{+34}_{-32}$ & $189^{+11}_{-17}$ & $188^{+27}_{-34}$ \\
  {Norm} & $5.8_{-0.5}^{+0.3}$ &$5.79_{-0.25}^{+0.4} $&$5.8_{-0.5}^{+0.4} $&$5.15_{-0.27}^{+0.4} $&$5.08_{-0.14}^{+0.12} $& $5.00_{-0.4}^{+0.18}$ \\
  \hline
  {\sc diskbb} && \\
  $T_{\rm in}$ [keV] & $0.437^{+0.018}_{-0.022}$ & $0.436^{+0.014}_{-0.009}$ & $0.437^{+0.018}_{-0.024}$ & $0.473^{+0.016}_{-0.018}$ & $0.446^{+0.017}_{-0.006}$ & $0.438^{+0.020}_{-0.03}$ \\
  Norm $[10^4]$ & $7.3_{-2.5}^{+3.8} $ &$7.4_{-5.2}^{+2.7} $ &$7.3_{-2.5}^{+3.8} $& $ 2.8_{-0.7}^{+0.6} $ & $ 4.52_{-0.19}^{+0.3} $ &$ 5.1_{-1.4}^{+3.6} $ \\
  \hline
  {\sc relconv\_nk} && \\
  $q_{\rm in}$ & $>7.6$ & $4.5_{-0.7}^{+1.7} $ & $>7.2$ & $>8.3$ & $8.8_{-2.0}^{+1.1}$ & $>8.0$ \\
  $q_{\rm out}$ & $3.4_{-0.3}^{+0.4} $ & $3.1_{-0.6}^{+0.3}$ & $3.43_{-0.23}^{+0.4} $ &$3.09_{-0.11}^{+0.12}$ & $2.88_{-0.07}^{+0.04}$ & $2.85_{-0.13}^{+0.08}$\\
  $R_{\rm br}$~$[M]$ & $2.67_{-0.27}^{+0.4} $ & $ 2.9_{-0.4}^{+3.1} $ & $2.7_{-0.3}^{+0.4} $ & $ 2.51_{-0.14}^{+0.18} $ & $ 1.55_{-0.09}^{+0.06} $ & $ 1.67_{-0.07}^{+0.11} $ \\
  $a_*$ & $ 0.957_{-0.011}^{+0.011} $ & $0.988_{-0.010}^{+0.004}$ & $>0.98$ & $0.948_{-0.011}^{+0.019}$ & $0.9914_{-0.008}^{+0.0012}$ & $>0.99$ \\
  $i$ [deg] & $43.5^{+2.2}_{-1.9}$ & $43.3^{+1.5}_{-1.0}$ & $43.6^{+2.3}_{-1.5}$ & $40.5^{+0.9}_{-1.0}$ & $40.6^{+0.6}_{-0.4}$ & $40.3^{+0.7}_{-0.8}$ \\
  $\alpha_{13}$ & $0^\star$ & $-0.77_{-0.08}^{+0.16} $ & $0^\star$ & $0^\star$ & $<-0.93$ & $0^\star$ \\
  $\alpha_{22}$ & $0^\star$ & $0^\star$ & $-0.19_{-0.04}^{+0.06} $ & $0^\star$ & $0^\star$ & $0.93_{-0.07}^{+0.06}$ \\
  \hline
  {\sc xillver} && \\
  $A_{\rm Fe}$ & $4.4^{+0.5}_{-0.4}$ & $4.4^{+0.5}_{-0.4}$ & $4.4^{+0.5}_{-0.4}$ & $3.95^{+0.4}_{-0.28}$ & $4.12^{+0.12}_{-0.28}$ & $4.2_{-0.4}^{+0.4}$ \\
  $\log\xi$ & $4.02^{+0.07}_{-0.05}$ & $4.03^{+0.05}_{-0.05}$ & $4.02^{+0.07}_{-0.04}$ & $3.84^{+0.05}_{-0.06}$ & $3.88^{+0.03}_{-0.04}$ & $3.92^{+0.08}_{-0.07}$ \\
  {Norm} & $0.131_{-0.020}^{+0.024} $ &$0.127_{-0.013}^{+0.003} $&$0.131_{-0.019}^{+0.023} $& $ 0.067_{-0.008}^{+0.009} $ & $0.071_{-0.003}^{+0.002} $& $ 0.073_{-0.005}^{+0.007} $ \\
  \hline
  {\sc gaussian} && \\
  {Norm} $[10^{-4}]$ &$6.2_{-2.3}^{+2.2} $ &$6.7_{-2.5}^{+2.1} $&$ 6.1_{-2.2}^{+2.2} $& $ 12.3_{-2.5}^{+2.0} $ &$13.1_{-2.5}^{+1.9} $& $ 12.5_{-2.9}^{+2.4} $ \\
  \hline
  $\chi^2/\nu$ & $1718.23/1603 \quad$ & $\quad 1716.28/1602 \quad$ & $\quad 1717.93/1602 \quad$ & $\quad 2083.58/1911 \quad$ & $\quad 2070.92/1910 \quad$ & $\quad 2078.26/1910 \quad$ \\
  & =1.07188 & =1.07134 & =1.07236 & =1.09031 & 1.08425 & 1.08810 \\
  \hline\hline
  \end{tabular}
  \vspace{0.2cm}
  \caption{Summary of the best-fit values assuming a broken power-law emissivity profile with both $q_{\rm in}$ and $ q_{\rm out} $ free. The reported uncertainties correspond to the 90\% confidence level for one relevant parameter. $^\star$ indicates that the parameter is frozen. \label{t-fit3}}
  \end{table*}

\section{Discussion \label{s-dis}}

Let us first consider the 6~Kerr fits, i.e.~3~models for the intensity profile for 2~spectra (epoch~1 and epoch~4). Generally speaking, our measurements are consistent with previous studies~\cite{c3,c4,c5,w16}, in particular we always recover a high spin parameter ($a_* \approx 0.95$) and an inclination angle $i \approx 40$~deg. This is quite independent of the intensity profile and the spectra. The fit with the intensity profile modeled by a simple power-law is surely worse than the fits with a broken power-law and we also find some difference in the inclination angle of epoch~1 and 4. When we assume a broken power-law, the quality of the fit improves and the estimate of the inclination angle between the two spectra becomes consistent. When $q_{\rm out}$ is free, the fit requires that its value is not far from 3, and indeed we do not improve the quality of the fit much with respect to the model with $q_{\rm out} = 3$. In Ref.~\cite{w16}, the authors report the best-fit values of their analysis assuming $q_{\rm out} = 3$ and their results can be directly compared with ours in Tab.~\ref{t-fit2}. All our measurements are consistent with their parameter estimates, with the exception of the column density and the outflow velocity for epoch~4. We do not know the origin of such a difference.

The situation for the 12~non-Kerr fits is more complicated. If we assume a simple power-law for the intensity profile, it seems that we cannot recover the Kerr solution; that is, the data require a non-vanishing deformation parameter at a significant confidence level (with the exception of epoch~4 with $\alpha_{22}$ free, where we recover Kerr when we consider the 90\% confidence level curve for two relevant parameters, see the top right panel in Fig.~\ref{f-c4}).

When we model the intensity profile with a broken power-law with $q_{\rm out} = 3$, the results change and now we can somehow recover the Kerr metric. The intensity profile thus matters here, while it seemed to be not so crucial in the estimate of the black hole spin under the assumption of the Kerr background. However, we clearly see that the $\chi^2$ minimizing algorithm of XSPEC has a problem to reliably find a minimum in such an extremely complicated $\chi^2$ landscape. MCMC simulations can better map such a complicated surface and this was indeed our main motivation to run them. The contour plot in the plane $a_*$ vs $\alpha_{13}$ of the MCMC simulations for epoch~1 shows larger uncertainties on $a_*$ and $\alpha_{13}$ and the spectrum is consistent with that expected in the Kerr metric (while there seems to be only marginal agreement in the plot obtained with the STEPPAR command in XSPEC in Fig.~\ref{f-c1}). A quick comparison of $\chi^2$ in the fits with a simple power-law and a broken power-law with $q_{\rm out} = 3$ clearly indicates that the discrepancy with the Kerr solution should not be as strong as it seems to be from the top panels in Figs.~\ref{f-c1} and \ref{f-c4}.

Lastly, we consider the fits with a broken power-law with $q_{\rm out}$ free. Since the best-fit of $q_{\rm out}$ is always close to 3, we should conclude that there are no substantial differences with the previous case with $q_{\rm out} = 3$. For the fits of epoch~1, this is indeed the case. For epoch~4, the bottom panels in Fig.~\ref{f-c4} suggest that we cannot recover the Kerr metric.


\section{Conclusions \label{s-con}}

In the present paper, we have shown our results of the analysis of two \textsl{NuSTAR} observations of Cygnus~X-1 in the soft state. Our constraints seem to be quite sensitive to the assumption of the emissivity profile. This was already argued in our previous work~\cite{noi7}, but here the problem is more severe, and the result is that we do not always recover the Kerr solution at an acceptable confidence level. This was not the case of our previous work, where the measurement of the deformation parameters was consistent with zero at 90\% confidence level and their uncertainty was much smaller (so we obtained strong constraints on the Kerr metric). There are presumably a few reasons: 
\begin{itemize}
\item The model employed here is more complex, in the sense that there are several components and thus many free parameters. This was not the case in other studies, see for instance the analysis of the bare active galactic nuclei reported in Ref.~\cite{noi9}, where the spectra are quite simple and we obtain very strong constraints on $\alpha_{13}$ and $\alpha_{22}$.
\item The absorption due to the wind of the massive companion star is modeled with {\sc xstar}, but inevitably introduces additional systematic uncertainties in the model that we would like to avoid in a test of the Kerr metric. A similar issue was probably in the analysis of GRS~1915+105 reported in~\cite{noi10}, where the outflow from the accretion disk limited the capability of measuring the spacetime metric.
\item Our non-relativistic reflection model {\sc xillver} is not appropriate for accretion disks of stellar-mass black holes. In {\sc xillver}, the calculations of the non-relativistic reflection spectrum are indeed done considering only the photons illuminating the disk by the corona and ignoring the thermal photons from the accretion disk itself. This is not a problem for supermassive black holes with temperatures of the inner part of the accretion disk in the range 1-100~eV, but it is for modeling the spectra of accretion disks around stellar-mass black holes, where the temperatures are in the soft X-ray band.  
\end{itemize}

To conclude, on the basis of the results in this work and in our previous papers, we can say that the choice of the right source and of the right observation is extremely important if we want to use X-ray reflection spectroscopy to test GR. Ideally, we can list the ``desired properties'' of source and observation as follows:
\begin{enumerate}
\item Supermassive black holes are probably more suitable than stellar-mass black holes. While stellar-mass black holes are typically brighter, their spectra are more difficult to model. The higher temperature of the accretion disk is one of the reasons.
\item We need fast-rotating objects ($a_* > 0.9$). This condition is necessary to break the parameter degeneracy. Intuitively, we can say that the reason is that the inner edge of the accretion disk is closer to the event horizon and relativistic effects are magnified.
\item No absorbers. Absorption material between the disk and the observer requires a model with some astrophysical uncertainties that we would like to avoid in a GR test. 
\item It is important to have data with a good energy resolution at the iron line and covering a broad energy band (for example, simultaneous observations by \textsl{XMM-Newton} and \textsl{NuSTAR}). The high energy resolution at the iron line is useful to resolve the relativistic features in the iron line. The broad energy band is necessary to break the parameter degeneracy and be able to estimate the cut-off energy (or the temperature) of the corona.
\item Prominent iron line. This follows from the fact the the iron line is the most informative feature for our tests of the Kerr metric.
\item Accretion luminosity between 5\% and 30\% of the Eddington limit. This is just the condition to have thin accretion disks~\cite{isco1,isco2}, but it is often violated in supermassive black holes. Currently we do not have an estimate of the impact of the thickness of the accretion disk and/or of the violation of the bound of 30\% of the Eddington limit, but work is underway. Since in our previous studies we got quite stringent constraints on the deformation parameters from supermassive black holes, and our measurements were always consistent with the Kerr metric at 1- or 2-$\sigma$, we may argue that current systematic uncertainties are subdominant, because a fine cancellation among them to recover the Kerr solution for all sources sounds unlikely, but we cannot say more as of now.
\item Corona with well-known geometry. The intensity profile of the reflection spectrum clearly plays an important rule in the measurements of the deformation parameters, and this will become surely more important with the next generation of X-ray missions and higher-quality data. An intensity profile described by a broken power-law is a crude approximation and does not permit to take all relativistic effects into account. In principle, the intensity profile could be theoretically predicted if we knew the exact geometry of the corona. It is likely that different coronal geometries are possible and even that the coronal geometry can change with time; for example, it is thought that the corona is extended (the accretion flow between a truncated disk and the black hole?) at the beginning of the hard state in stellar-mass black holes and becomes compact (the base of the jet?) later as the outburst evolves (see, for instance, \cite{corona-wp,book} and references therein). It would be helpful to test black holes with a well understood coronal geometry, so that the intensity profile can be predicted theoretically.
\end{enumerate}

The conclusion of the present work is that GR tests require suitable sources. Cygnus~X-1 seems to have a spectrum that is too complicated for our purpose and our current modeling capabilities, and the supermassive black holes studied in our previous papers are definitively more promising candidates. It is difficult to generalize our conclusion to all stellar-mass black holes. A fundamental limit is that there are not so many sources available that we can select only those with all the desirable properties listed above. If we requires $a_* > 0.9$ (point~2), we only a know a few stellar-mass black holes that may have a high value of the spin parameter, while most supermassive black holes with a spin estimate meet this condition (say, around 20 sources).

Precision tests of GR using X-ray reflection spectroscopy may be possible in the future if we have sufficiently sophisticated models to describe every component of a source. If we assume that the deformation parameters have the same values for every black hole (as it is the case in many modified theories of gravity), we do not need to be able to model all sources. We can just focus the attention on a specific source (or a few specific sources) with good properties for tests of the Kerr metric, a well-understood coronal geometry, and high-quality data and test all possible deviations from Kerr there.


{\bf Acknowledgments --}
We thank Dominic Walton for providing us his {\sc xstar} table. This work was supported by the Innovation Program of the Shanghai Municipal Education Commission, Grant No.~2019-01-07-00-07-E00035, and Fudan University, Grant No.~IDH1512060. A.B.A. also acknowledges the support from the Shanghai Government Scholarship (SGS). J.A.G. and S.N. acknowledge support from the Alexander von Humboldt Foundation. S.N. also acknowledges support from the Excellence Initiative at Eberhard-Karls Universit\"at T\"ubingen.


\end{document}